\documentclass[a4paper,11pt]{article}

\usepackage[normalem]{ulem}
\usepackage{jheppub} 
\pdfoutput=1
\usepackage{graphicx,array}
\usepackage[dvipsnames]{xcolor}
\usepackage{amstext}
\usepackage{amssymb}
\usepackage{slashed}
\usepackage{cancel}
\usepackage{subcaption}
\usepackage{booktabs}
\usepackage[skins,theorems]{tcolorbox}
\usepackage{tikz}
\usepackage{tikz-3dplot}
\usepackage{multirow}
\usetikzlibrary{calc}
\usetikzlibrary{decorations} %
\usepackage[UKenglish]{babel}
\usepackage[toc,page]{appendix}
\usepackage{amsmath}
\usepackage{amssymb}
\usepackage{graphicx}
\usepackage{caption}
\usepackage[vcentermath]{youngtab}
\usepackage{slashed}
\usepackage{color}
\usepackage{stackrel}
\usepackage{tikz-cd} 
\usepackage{cancel} 

\usetikzlibrary{shapes,matrix}

\newenvironment{eqn*}{\begin{equation*}\begin{aligned}}{\end{aligned}\end{equation*}\noindent}

\newcommand{\bqa}{\begin{eqnarray}}
\newcommand{\eqa}{\end{eqnarray}}

\newtheorem{theorem}{Theorem}
\newtheorem{corollary}[theorem]{Corollary}

\newtheorem{proposition}{Proposition}

\newcommand{\be}{\begin{equation}}
\newcommand{\ee}{\end{equation}}
\newcommand{\beq}{\begin{equation}}
\newcommand{\eeq}{\end{equation}}
\newcommand{\ba}{\begin{aligned}}
\newcommand{\ea}{\end{aligned}}

\newcommand{\bea}{\begin{eqnarray}}
\newcommand{\eea}{\end{eqnarray}}

\newcommand{\cT}{\mathcal{T}}

\newcommand{\cL}{\mathcal{L}}

\newcommand{\cN}{\mathcal{N}}

\newcommand{\cA}{\mathcal{A}}

\newcommand{\cF}{\mathcal{F}}

\newcommand{\cV}{\mathcal{V}}

\newcommand\bi{\begin{itemize}}
\newcommand\ei{\end{itemize}}

\def\unit{{1\kern-.65ex {\rm l}}}
\def\1{{1\kern-.65ex {\rm l}}}

\newcount\hour \newcount\minute
\hour=\time \divide \hour by 60
\minute=\time
\def\now{%
\ifnum \hour<13
  \ifnum \hour=0 \advance \hour by 12 \number\hour:\else \number\hour:\fi%
     \ifnum \minute<10 0\fi%
     \number\minute%
\ A.M.%
\else \advance \hour by -12 \number\hour:%
  \ifnum \minute<10 0\fi%
  \number\minute%
  \ P.M.%
\fi%
}

\makeatother

\title{Density of States, Black Holes and the Emergent String Conjecture}
\author[a,b]{Alek Bedroya}
\author[b]{, Rashmish K.~Mishra}
\author[b]{, and Max Wiesner}
\affiliation [a]{Princeton Gravity Initiative, Princeton University, Princeton, NJ 08544, USA}
\affiliation[b]{Jefferson Physical Laboratory, Harvard University, Cambridge, MA 02138, USA}

\abstract{
We study universal features of the density of one-particle states $\rho(E)$ in weakly coupled theories of gravity at energies above the quantum gravity cutoff $\Lambda$, defined as the scale suppressing higher-derivative corrections to the Einstein--Hilbert action. Using thermodynamic properties of black holes, we show that in asymptotically flat spacetimes, certain features of $\rho(E)$ above the black hole threshold $M_{\rm min}$ are an indicator for the existence of large extra dimensions, and cannot be reproduced by any lower-dimensional field theory with finitely many fields satisfying the weak energy condition. Based on the properties of gravitational scattering amplitudes, we argue that there needs to exist a (possibly higher-dimensional) effective description of gravity valid up to the cutoff $\Lambda$. Combining this with thermodynamic arguments we demonstrate that $\rho(E)$ has to grow exponentially for energies $\Lambda \ll E \ll M_{\rm min}$. Furthermore we show that the tension of any weakly coupled $p$-brane with $p\geq 1$ is bounded from below by $\Lambda^{p+1}$. We use this to argue that any tower of weakly coupled states with mass below $\Lambda$ has to be a Kaluza--Klein (KK) tower. Altogether these results indicate that in gravitational weak-coupling limits the lightest tower of states is either a KK tower, or has an exponentially growing degeneracy thereby resembling a string tower. This provides evidence for the Emergent String Conjecture without explicitly relying on string theory or supersymmetry.}
\begin{document}
\maketitle
\flushbottom

\section{Introduction and Summary}

A characteristic feature of gravitational theories is the existence of massive states beyond those describable by an effective field theory (EFT). Such states could be strongly coupled (e.g. black hole microstates) or weakly coupled (e.g. string excitations). Oftentimes, these states provide crucial information about the theory of gravity and are believed to be required for consistency of the theory at the quantum level. Whereas the exact spectrum of the massive states depends on the details of the UV completion, one may expect it to have some features that are universal for any theory of gravity. Proposals for such universal features of the spectrum of the massive states in quantum gravity have been formulated in the context of the Swampland program~\cite{Vafa:2005ui}, e.g., through the Distance Conjecture~\cite{Ooguri:2006in}. These proposals are based on evidence from string theory and have been tested to a great extent in a plethora of explicit string compactifications (see \cite{Palti:2019pca,vanBeest:2021lhn,Agmon:2022thq} for reviews). 
Whereas there is strong evidence for such conjectures in the context of string theory, a pressing question concerns their validity in general theories of quantum gravity away from the string theory lamppost. 

In this spirit, the goal of this work is to extract general properties of the spectrum of massive states in quantum gravity regardless of the details of the UV completion. The central object of our analysis is the density of one-particle states, $\rho$, in quantum gravity. In particular we are interested in extracting the universal behavior of $\rho$ as a function of energy $E$.\footnote{For one-particle states we refer to the rest mass $m$ as ``energy'' in the following.} Concretely, we aim to study the properties of $\rho(E)$ at energies above the quantum gravity cutoff. Combining results from gravitational scattering amplitudes with thermodynamics for weakly coupled theories of gravity, we relate the high-energy behavior of $\rho(E)$ to the properties of towers of states below the black hole thresholds. This, in turn, provides crucial insight into the general properties of weakly coupled theories of gravity.

There are multiple notable energy scales that are relevant for the study of high energy states in quantum gravity. The first energy scale is the Planck scale $M_{\rm pl}$ which describes the strength of the gravitational coupling. Another energy scale is the scale at which gravity becomes strongly coupled. This scale is also known as the species scale~\cite{Dvali:2007hz,Dvali:2007wp,Dvali:2009ks,Dvali:2010vm,Dvali:2012uq} which, in theories with many weakly coupled particles, can be parametrically below the Planck scale, $\Lambda\ll M_{\rm pl}$. A precise proposal to define the species scale has been given in~\cite{vandeHeisteeg:2022btw,vandeHeisteeg:2023ubh} that identifies the species scale with the scale that suppresses gravitational higher-derivative corrections to the Einstein--Hilbert action. Indeed, the higher-derivative corrections have been confirmed to be a good measure for the species scale in various explicit string theory setups~\cite{Cribiori:2022cho,vandeHeisteeg:2023uxj,Cribiori:2023sch,vandeHeisteeg:2023dlw,Castellano:2023aum}.\footnote{See \cite{Calderon-Infante:2023uhz} for a discussion of the equivalence of the various definitions of the species scale in explicit string theory setups.}

It is often stated that the species scale, defined as the scale suppressing the higher-derivative corrections, is also the curvature scale at the horizon of the smallest black hole describable within effective field theory~\cite{Dvali:2007hz,Dvali:2007wp,vandeHeisteeg:2022btw}. However, the inverse radius of the smallest black hole describable in a given EFT can be parametrically different from the species scale. The scale associated to the inverse radius of this smallest black hole was recently conjectured in~\cite{Bedroya:2024uva} to be a universal scale in quantum gravity, called the black hole scale $\Lambda_{\text{BH}}$. It was also suggested that $\Lambda_{\rm BH}$---which is defined using black hole physics---is a well-defined substitute for the mass scale of the lightest weakly coupled tower of particles at low-energies. Let us briefly summarize this main observation of~\cite{Bedroya:2024uva}. Consider the low-energy action of a theory given by the dimensional reduction of a higher-dimensional theory on a compact manifold of large radius $R$. 
In case the species scale of the higher-dimensional theory is the higher-dimensional Planck scale, this scale will also be the species scale of the lower-dimensional theory. If the radius of the horizon of the smallest black hole describable in the lower-dimensional effective field theory was set by the inverse species scale, one would conclude that any lower-dimensional black hole solution with horizon size larger than the higher-dimensional Planck scale is reliable. However, due to the Gregory--Laflamme instability~\cite{Gregory:1993vy}, a black hole for which the length scale of the horizon is much smaller than the radius, $R$, of the compact space is unstable and decays into higher-dimensional black holes which are localized in the extra-dimension. Therefore, the smallest black hole that can be described by the lower-dimensional EFT has inverse-radius $\Lambda_{\rm BH}\sim 1/R$.

The above argument demonstrates two points: (1) There is often a rich hierarchy of scales in quantum gravity ($\Lambda_{\rm BH}\ll \Lambda_s\ll M_{\rm pl}$) which leaves an imprint both at low energies (existence of particle towers) as well as high energies (Gregory--Laflamme transition). (2)~The connection between the higher-derivative corrections and the size of the smallest black hole is more subtle than previously thought. In fact, the hierarchy of scales could be extended to include more energy scales by having multiple Gregory--Laflamme transitions in the presence of extra dimensions with parametrically different length scales. In this work, we provide a unifying framework to study the different energy scales and their connections to one another by studying the behavior of $\rho(E)$ across a wide range of energies. In particular, we shed light on the connection between the species scale and the \textit{smallest black hole}. The key point is to clarify what is meant by the smallest black hole. For example if, in the presence of large extra dimensions, ``smallest black hole'' refers to the smallest higher-dimensional black hole, the radius of this black hole is indeed expected to be $\mathcal{O}(\Lambda^{-1})$. In this paper we show that the smallest black hole in this more general sense can be probed by gravitational amplitudes and demonstrate how its size is related to the higher-curvature expansion of the effective action. In that way we settle the problem of relating the size of the \textit{smallest} black holes to the species scale defined in terms of the higher-curvature corrections. Our results can be summarized in the following way: the inverse of the horizon size of the smallest black hole describable by any EFT is given by the cutoff of this EFT. Importantly this EFT cutoff can differ from the energy scale controlling the higher-curvature corrections. In particular if the theory arises as compactification of a higher-dimensional theory, the cutoff of the higher- and lower-dimensional EFTs are different. The lower-dimensional theory has a tower of KK modes with mass $m\ll \Lambda$ that can never be included in an EFT in a parametrically controlled way. Therefore, the cutoff of the lower-dimensional EFT is given by the mass scale of the KK tower $\sim 1/R$. The lower-dimensional EFT describes black holes with horizon radius $r_H\gg R$, while in the higher-dimensional EFT it is possible to describe much smaller black holes with radius $R>r_H\gg 1/\Lambda$.

Since our arguments rely on the dependence of the density of one-particle states $\rho$ on the energy, it is crucial to have a reliable way of calculating $\rho$ at high energies above the EFT cutoff. By studying the relation between scattering amplitudes and the density of one-particle states for black holes and strings, we propose an approximate measure for $\rho(E)$ at energies $E\gg\Lambda$ in terms of gravitational scattering amplitudes. Given this estimate of $\rho(E)$ we study its behavior for high energies which we utilize to infer the structure of possible towers of states in quantum gravity with mass below the black hole threshold. To that end, we work in asymptotically flat spacetimes and focus on certain gravitational weak-coupling limits. These are limits in the scalar field space of gravitational theories where the gravitational backreaction vanishes for massive particles with fixed mass (in units of $\Lambda$) below the black hole threshold. Particles with vanishing gravitational backreaction are to be contrasted to large black holes whose mass is entirely sourced by their gravitational self-energy. For states with fixed mass (in units of $\Lambda$) to have vanishing gravitational backreaction, we require $\Lambda \ll M_\text{pl}$. Therefore for gravitational weak-coupling limits the quantum gravity cutoff, or species scale, can be made parametrically smaller than the Planck scale, i.e., $\Lambda\ll M_{\rm pl}$. Since the species scale relates to the number of light species in the theory, such limits feature towers of light states. Hence gravitational weak-coupling limits are exactly those kinds of limits in moduli space that are of interest in the context of the Distance Conjecture. 

The nature of the possible towers of states in such infinite distance limits is believed to be strongly constrained by the Emergent String Conjecture (ESC)~\cite{Lee:2019oct}. This conjecture states that, at least in asymptotically flat spacetime, any infinite distance limit of a theory of gravity is either a limit in which a fundamental string becomes tensionless or a decompactification limit to a higher-dimensional theory. According to the ESC the lightest tower of states becoming asymptotically massless at an infinite distance in the scalar field space either corresponds to a KK-tower or the excitation tower of a light string. In particular, the ESC prohibits the lightest tower to arise from membranes or other extended objects, see \cite{Alvarez-Garcia:2021pxo} for a discussion on this in five-dimensional compactifications of M-theory. The ESC has been tested and confirmed in different string theory setups \cite{Lee:2018urn,Lee:2018spm,Lee:2019jan,Lee:2019apr,Baume:2019sry,Klaewer:2020lfg, Lee:2021usk,Wiesner:2022qys,Bedroya:2023tch,Etheredge:2023odp,Alvarez-Garcia:2023qqj} and has proven to be a powerful ingredient for many recent bottom-up arguments (see e.g. \cite{Lanza:2021udy,Rudelius:2021oaz,Etheredge:2022opl,Montero:2022prj,vandeHeisteeg:2022btw,Rudelius:2022gbz,vandeHeisteeg:2023ubh,Bedroya:2023xue,Bedroya:2023tch}). Still, a definite bottom-up proof for the stringent constraints on the possible nature of the lightest tower of states predicted by the ESC is lacking---even though~\cite{Basile:2023blg} presented arguments in its favor based on the concept of species scale thermodynamics \cite{Cribiori:2023ffn,Basile:2024dqq}. In this work we take a different approach and use the properties of the density of one-particle state and gravitational scattering amplitudes to argue that the lightest tower of states in gravitational weak-coupling limits are either KK-towers or a tower of states with degeneracy growing exponentially fast in energy thereby resembling the excitation tower of a fundamental string. Let us stress, however, that we do not identify the origin of this tower of states with exponential degeneracy, and in particular cannot answer whether or not they arise from the excitations of a fundamental string. Still, we view our results as string theory independent evidence for the ESC.

This paper is structured as follows: the rest of this section provides an overview of the main results of the paper. We review the definition of~\cite{vandeHeisteeg:2023dlw,vandeHeisteeg:2023ubh} of the quantum gravity cutoff in terms of higher-derivative corrections to the Einstein--Hilbert action in section~\ref{ssec:speciesscale}. In section~\ref{sec:density} we introduce a way to estimate the density, $\rho$ of one-particle states in terms of high-energy scattering amplitudes and discuss some basic properties of the dependence of $\rho$ on $E$. In section~\ref{spscandamplitudes} we use gravitational scattering amplitudes to show the equivalence between the cutoff of the gravitational EFT describing \emph{all} black holes in the theory and the species scale, and present a bound on the tension of weakly coupled $p$-branes in terms of the species scale. Section~\ref{sec:spectrum} studies the spectrum of weakly coupled gravitational theories via the energy dependence of $\rho$ at energies above the species scale which enables us to connect our results to the ESC. We present our conclusions and future perspectives in section~\ref{sec:discussion}. The appendices contain some background on effective strings, Gross--Mende saddles and scattering amplitudes that are the basis of the arguments presented in the main text. 

\subsection{Summary of Results}
The central objects of our analysis are the species scale $\Lambda$ and the density of one-particle states $\rho(E)$. Motivated by the behavior of scattering amplitudes in the black hole and perturbative string regimes at energies above the cutoff, $E>\Lambda$, we define an approximate notion for $\rho$ in terms of $2\to 2$ scattering amplitudes at fixed impact parameter $b\lesssim \Lambda^{-1}$. The precise identification is given in \eqref{ADOS}. The energy dependence of $\rho$ can be parametrized as
\begin{equation}
    \rho(E) \sim \exp\left[\left(\frac{E}{\Lambda}\right)^\alpha\right]\,,
\end{equation}
for some $\alpha>0$ up to terms polynomial in $E$. The parameter $\alpha$ is understood to be a piece-wise constant function of the energy $E$. As we show, the value of $\alpha$ in a given energy window encodes crucial information about the nature of the one-particle states with mass in that energy window. For example, using basic properties of black hole thermodynamics we show that in case $\alpha>1$ for some energy $E_0$, the corresponding one-particle states with mass $m=E_0$ are necessarily black hole microstates, see Proposition~\ref{propparticleBH}. In other words, the energy $M_{\rm min}$ corresponding to the transition between particles and black holes can be detected in the density of one-particle states as the energy at which $\alpha\leq 1$ changes to $\alpha>1$. The energy scale $M_{\rm min}$ then corresponds to the minimal mass of a black hole describable in the effective theory. Let us stress that this black hole may be a higher-dimensional black hole, in case the effective theory arises as a compactification of a higher-dimensional effective theory of gravity.

The curvature at the horizon of the smallest black hole describable in the (possibly higher-dimensional) EFT sets a scale $\Lambda_{\rm min}$ that serves as cutoff of the EFT. Still, it is not obvious that in general this scale should be identified with the scale $\Lambda$ appearing in the higher-derivative corrections to the Einstein--Hilbert action. Let us note, however, that in \cite{Cribiori:2022nke} the equivalence of these two scales has been demonstrated in the specific case of supersymmetric black holes in four-dimensional $\cN=2$ theories. In this work we argue for the identification of these two scales by using gravitational scattering amplitudes. To that end, we focus on $2\to2$ graviton scattering in the unphysical regime in which the Mandelstam variable $t\gg 0$. To argue for $\Lambda=\Lambda_{\rm min}$ we use that in the unphysical regime and for center of mass energies above the black hole threshold the amplitude grows exponentially in the radius of a black hole of that mass. We then show that for energies $\Lambda_{\rm min}\ll \sqrt{s}\ll M_{\rm min}$ this radius has to be replaced by $\Lambda_{\rm min}^{-1}$, i.e. the radius of the smallest black hole describable in the theory. We use this amplitude to match the coefficients of the higher-derivative expansion using the  fact that below $\Lambda_{\rm min}$ the full effective action including all higher-derivative terms has to give the full amplitude. This leads to the identification $\Lambda_{\rm min}=\Lambda$ as stated in Proposition~\ref{prop:existenceofEFT} implying the existence of a (possibly higher-dimensional) EFT description all the way up to $\Lambda$. To the best of our knowledge, this provides the first general gravitational argument as to why the scale suppressing the higher-derivative corrections is the same as the scale of the curvature at the horizon of the smallest black hole describable by \textit{some} effective theory. 

We further use gravitational scattering amplitudes to impose a bound on the tension of weakly coupled strings in quantum gravity. Since these strings are weakly coupled we argue that their worldsheet theory can be well-described in terms of the Nambu--Goto action using the effective string theory of Polchinski and Strominger~\cite{Polchinski:1991ax} --- even if the string is not a critical string. The contribution of the Gross--Mende saddles~\cite{Gross:1987ar} of these weakly coupled strings to the scattering amplitude can then be compared to the scattering amplitude expected from the full effective action including higher-derivative terms. Using this comparison, we infer that the tension $\cT$ of any weakly coupled string has to be bounded as 
\begin{equation}\label{boundstring}
    \cT\gtrsim \Lambda^2\,. 
\end{equation}
Since a weakly coupled $p$-brane with $p>1$ yields a weakly coupled effective string upon dimensional reduction we can extend this bound to general $p$-branes as claimed in Proposition~\ref{prop:tension}. Notice that, for the specific case of 4d $\cN=1$ theories gravity, \cite{Martucci:2024trp} argued for a bound on  the tension of weakly coupled strings which is precisely of the form~\eqref{boundstring} and hence consistent with our general argument.

Using this bound and the fact that there exists a (higher-dimensional) EFT description for the theory valid up to the energy $\Lambda$ we then conclude that in gravitational weak-coupling limits $\Lambda\ll M_{\rm pl,d}$ any tower of states with mass below the species scale has to be a KK tower. This is summarized in Proposition~\ref{prop:lighttower}. For energies above the species scale but below the black hole threshold, i.e. for $\Lambda\ll E\ll M_{\min}$, we employ basic thermodynamic arguments to show that there exists a tower of one-particle states with mass proportional to $\Lambda$ whose degeneracy grows exponentially in the energy. In other words the Hagedorn behavior
\begin{equation}
    \rho(E) \propto \exp\left(\frac{E}{\Lambda}\right)\,,\quad \text{for}\quad \Lambda\ll E \ll M_{\rm min}\,,
\end{equation}
is a universal property of the spectrum of quantum gravity as stated in Proposition~\ref{prop:Hagedorn}. Finally, we present an argument that indicates that above the black hole threshold we can use the scaling of $\rho$ as a function of energy to detect the existence of extra dimensions and their size by showing that (higher-dimensional) Schwarzschild black hole are always the dominant contribution to $\rho(E)$ and uniquely determine $\alpha$ in a given spacetime dimension as stated in Proposition~\ref{prop:higherdimBH}. 

Using our results on the possible tower of states in gravitational weak-coupling limits, we conclude in Corollary~\ref{mESC} that any tower in a weakly coupled theory of gravity is either a KK tower or exhibits Hagedorn behavior which we view as evidence for the key statement of the ESC that the lightest tower of states is either a KK tower or the excitation tower of a weakly coupled string.

\section{Review: Species scale}\label{ssec:speciesscale}
Our analysis crucially relies on two quantities that we can define in theories of quantum gravity: the quantum gravity cutoff (or species scale) $\Lambda$ and the density, $\rho$, of one-particle states. In this section we review the definition of the species scale in terms of higher-derivative operators as originally proposed in~\cite{vandeHeisteeg:2022btw,vandeHeisteeg:2023dlw,vandeHeisteeg:2023ubh}. The relation of the species scale defined in this way to the size of the smallest black hole will be made precise in section~\ref{sec:ExistenceEFT}.

To define the species scale we start with a $d$-dimensional theory of gravity that at low energies can be described by Einstein gravity with action 
\begin{equation}\label{SSD}
    S_{\rm EH} = \frac{M_{\rm pl,d}^{d-2}}{2} \int {\rm d}^d x \sqrt{-g} \left(\mathcal{R} + \cdots \right)\,,
\end{equation}
where $M_{\rm pl,d}$ is the $d$-dimensional reduced Planck scale, $g$ is the metric on the $d$-dimensional spacetime, $\mathcal{R}$ the Ricci scalar and the $\cdots$ stand for terms involving additional fields coupled to gravity. Radiative corrections to the effective action of the quantum theory are expected to generate an infinite series of gravitational higher-derivative operators involving the Riemann tensor and derivatives thereof. In other words, the Einstein--Hilbert action above will be corrected by terms of the form 
\begin{equation}\label{scorr}
    S_{\rm corr.} = \frac{M_{\rm pl,d}^{d-2}}{2} \int {\rm d}^d x \sqrt{-g} \left(\sum_{n=1}^{\infty} a_n \frac{\mathcal{O}_{2n+2}(\mathcal{R})}{M_{\rm pl,d}^{2n}}\right)\,. 
\end{equation}
Here, $\mathcal{O}_m(\mathcal{R})$ are dimension-$m$ operators involving the Riemann tensor $\mathcal{R}$ and its derivatives such as $\mathcal{R}^2$ and $\mathcal{R}\,\Box \mathcal{R}$ which are examples for dimension-4 and -6 operators, respectively. The coefficients $a_n$ are the Wilson coefficients of the respective operators that encode the imprints of quantum gravity on the effective action. The coefficients $a_n$ in general depend on $n$ and, in the presence of light scalar fields $\phi$, also on the vevs for these scalar fields~\cite{vandeHeisteeg:2022btw,vandeHeisteeg:2023ubh}. There typically exist limits in the scalar field space of the theory of quantum gravity in which the quantum gravity cutoff $\Lambda$ differs significantly from the Planck scale. This manifests itself in the Wilson coefficient $a_n(\phi)$ being parametrically large. Given a higher-derivative term one then expects the relation 
\begin{equation}\label{anLambda}
    \frac{\Lambda(\phi)}{M_{\rm pl,d}} \sim a_n(\phi)^{-1/2n}\,,
\end{equation}
In principle there can be some generic suppression of the Wilson coefficients with $n$. For the string case, we discuss this $n$-dependence in more detail around \eqref{hatan}. Taking these $n$-dependent prefactors into account we conclude that $\Lambda$ indeed replaces $M_{\rm pl,d}$ as cutoff scale in the effective action~\eqref{SSD}. Notice that in practice this could give a slightly different species scale depending on which term in the effective action is considered. To avoid this ambiguity~\cite{vandeHeisteeg:2023dlw} proposed to define the quantum gravity cutoff as
\begin{equation}\label{deflambda}
    \Lambda(\phi) \sim \frac{M_{\rm pl,d}}{\sup_{n\geq 1} \left\{a_n(\phi)^{\frac{1}{2n}}\right\}}\,. 
\end{equation}
For self-consistency the Wilson coefficients have to be evaluated at the scale on LHS of the above equation. 

Since the Wilson coefficient in the effective action \eqref{scorr} are energy-dependent, we have to choose an appropriate energy scale $\mu$ to define $\Lambda$. If we push $\mu$ to higher energies, we must integrate in all particles with mass below $\mu$ into the EFT. However, this process cannot be continued indefinitely due to the existence of black holes. In other words, there is an energy threshold beyond which the particles do not admit a field theory description. The energy scale $\mu$ at which we define $\Lambda$ is given by the inverse radius of the smallest (potentially higher-dimensional) black hole in the theory. This scale is a natural UV cutoff in quantum gravity, see \cite{Brustein:2009ex} for a discussion on this.
Equivalently we should consider the EFT in which all black holes in the theory can be described. Only in this case does the definition in \eqref{deflambda} give the correct quantum gravity cutoff. To illustrate this point let us, for the moment, consider a low-energy EFT in which a light state with mass $m_0$ has been integrated out. From standard EFT expectation, the Wilson coefficients
of a general higher-derivative operator of dimension $2n+2$ scale as 
\begin{equation}
    a_n \sim \left(\frac{M_{\rm pl,d}}{m_0}\right)^{2n+2-d}\,.% \qquad \Rightarrow \qquad a_n^{-1/2n} \sim \left(\frac{M_{\rm pl,d}}{m_0}\right)^{-\frac{2n+2-d}{2n}}\,. 
\end{equation}
A low-energy observer unaware of the existence of the particle with mass $m_0$ who measures the Wilson coefficient of the gravitational higher-derivative operators for large enough $n$ and applies \eqref{deflambda} would now erroneously conclude that the species scale is set by $m_0$. However, if $m_0\ll M_{\rm pl,d}$ the scaling of the higher-derivative Wilson coefficients with $n$ is very different from what we expect for the actual species scale. In the latter case, the value of $a_n^{-1/2n}$ for different $n$ only differs by the aforementioned $n$-dependent constants that will be discussed in section~\ref{sec:ExistenceEFT}. In contrast, in case the $a_n$ are determined by the light state with mass $m_0\ll M_{\rm pl,d}$, the coefficients $a_n^{-1/2n}$ differ by powers of $M_{\rm pl,d}/m_0\gg1$. The $n$-dependence of $a_n^{-1/2n}$ is hence an indicator of whether the low-energy effective action misses some light particles. The Wilson coefficients $a_n$ only give the correct species scale for the EFT in which all light particles have been integrated in. 

As argued in \cite{Dvali:2007hz,Dvali:2007wp,Dvali:2010vm,Dvali:2012uq}, the effective quantum gravity cutoff is expected to be significantly lowered compared to the Planck scale in the presence of many species of light particles. In this context, the quantum gravity cutoff is referred to as the species scale as it effectively counts the number of light species in the theory of gravity. According to the distance conjecture~\cite{Ooguri:2006in} large numbers of light species arise in asymptotic, infinite distance regimes in the moduli space of gravitational theories. In these regimes the parametric scaling of the species scale can then be extracted from the properties of the tower of light states. As shown for concrete examples in~\cite{vandeHeisteeg:2022btw,vandeHeisteeg:2023dlw} the scaling of the species scale in these limits is exactly reproduced by the coefficients of the higher-derivative corrections. Therefore \eqref{deflambda} indeed provides a definition of the species scale, $\Lambda$, which furthermore is also valid in the interior of the moduli space. 
In our work, we follow this proposal to define the species scale as it is more suitable for our approach since it does not directly refer to the number of light species. 

As we stress in section~\ref{sec:ExistenceEFT}, the fact that the species scale is identified with the overall scale suppressing the higher-derivative expansion does, however, not necessarily mean that the higher-derivative expansion diverges for curvature scale of order $\Lambda^2$. Instead, the appearance of the species scale in the full effective action
\begin{equation}
    S_{\rm eff.} = S_{\rm EH} + S_{\rm corr.}\,,
\end{equation} 
is required so that it correctly computes certain high energy ($E\gg\Lambda$) scattering amplitudes in the full theory of gravity.

\section{Density of one-particle states}\label{sec:density}
Defining a notion of the density of one-particle states in theories of quantum gravity is a challenging task. Whereas for energies below the cutoff (i.e. the species scale $\Lambda$), one can just count the number of states in the EFT, it is not clear how to do this at energies above $\Lambda$. In section~\ref{OPS}, we discuss some challenges in defining the one-particle density of states and provide a practical way to approximate the exact $\rho(E)$ at high energies that addresses these issues. Basic properties of $\rho$ as a function of energy are discussed in section~\ref{ssec:rhoofE} which are utilized in section~\ref{sec:particleBH} to infer the energy of the particle-black hole transition from~$\rho(E)$. 
\subsection{Defining \texorpdfstring{$\rho(E)$}{}: Challenges and Resolution}\label{OPS}
The analysis in this paper crucially relies on properties of the density of one-particle states $\rho$ as a function of energy $E$. For a one-particle state we consider its energy to be given by its rest mass $m$ and in this work we use energy and mass interchangeably unless the distinction is important. For energies below the cutoff, $E\ll \Lambda$, the density of states can be determined simply by counting the states in the field theory. On the other hand at energies $E\gg \Lambda$ computing the density of one-particle states is more difficult due to some subtleties arising in defining an exact notion of $\rho(E)$. Before we provide a practical way how to estimate $\rho(E)$ let us briefly comment on these issues:

\begin{enumerate}
    \item \textbf{Unstable particles:} Unstable particles can contribute as intermediate states in a scattering process, but do not show up as asymptotic states. Depending on the decay rate $\Gamma$ of the unstable particles (in units of their mass $M$), they may need to be included in the density of states to correctly account for the thermodynamic properties. For example, consider systems with a sufficiently long lifetime that can nevertheless be treated thermodynamically in the adiabatic approximation (e.g. large black holes in flat space). For such systems, unstable particles with sufficiently long lifetime are thermodynamically relevant. However, this makes it challenging to define the density of states in a precise way that includes such sufficiently long-lived unstable particles.
    %
    %
    %It is challenging to give a precise definition of the mass of an unstable particle in quantum gravity. While unstable particles lead to a pole in the local correlation functions, any attempt to define local observables in quantum gravity is rendered imprecise by exponentially small effects mediated by black holes~\cite{Arkani-Hamed:2017jhn}. Alternatively, the fundamental observables in quantum gravity are believed to have a precise definition on the boundary of space time. In Minkowski space, which is the focus of this work, these boundary observables take the form of scattering cross-sections. However, unstable particles eventually decay and cannot be asymptotic states. Therefore, their definition must rely on the scattering amplitude of stable particles. If one analytically continues the scattering amplitude in terms of the Mandelstam variable $s$, it has a branch-cut at the threshold of creating multi-particle states. This branch cut leads to different Riemann sheets for the scattering amplitude. One of these sheets is the \textit{physical sheet} which is used for calculating physical cross-sections. It is well-known that unstable particles do not lead to any poles on the physical sheet. However, the perturbative calculation would lead to a pole on the first unphysical Riemann sheet~\cite{Gunson:1960fha}. Identifying the mass of the unstable particle via this pole relies on the analytic continuation of the scattering amplitude, which itself must be derived from the physically observed cross-section. 
    %
    \item \textbf{Bound states:} A bound state is a state in the multi-particle Hilbert space of stable particles which is captured in the density of states. We are interested in using $\rho(E)$ to estimate the entropy, by mapping it to a system of free particles. This raises the question whether including the bound states will over-count the entropy. We note that the free particle approximation is reasonable as long as the particles are sufficiently far from each other. Consider a system with particles of type $p$, that can form a two-particles bound state $b$, with size $r_b$. To calculate the entropy, we approximate a state of the original system consisting of particles of type $p$ with a collection of particles of type $p$ and $b$, that are separated by distances larger than $r_b$. The inclusion of bound states is needed for a free-particle counting to correctly capture the entropy of an interacting system. 
    \item \textbf{Mass averaging:} Finally, even if we could define the mass of unstable particles in a sharp way, the density of one-particle states would still be a collection of $\delta$-functions. In this work we are, however, interested in studying the behavior of some continuous function $\rho(E)$, i.e., an averaged version of the one-particle density. 
\end{enumerate}

Given these challenges in defining $\rho(E)$ we now provide a way to estimate the actual density of one-particle states. This estimate has to address the points raised above and should correctly approximate the number of black hole microstates at energies above the black hole threshold. As we argue below for energies $E\gg \Lambda$ a good estimate of $\rho(E)$ is given in terms of the $2\to2$ scattering amplitude $\mathcal{A}_{2\to 2}(E)$ at fixed small impact parameter $b\lesssim \Lambda^{-1}$ as
\begin{equation}\label{ADOS}
    -\log \left(\Lambda\rho(E)\right)\sim \log \left|\mathcal{A}_{2\to 2}(E)\right|^2 + \mathcal{O}(\log (\Lambda/E))\,.
\end{equation}
Since $\rho$ is the density of states per unit energy, it has inverse mass dimension. As indicated in the above expression, in this work we consider $\rho$ in units of the inverse quantum gravity cutoff which in the following we assume implicitly and drop the $\Lambda$ factor in front of $\rho(E)$.

Let us now explain why \eqref{ADOS} is a reasonable estimate of the actual $\rho(E)$: At high energies and low-impact parameters, the in-going particles have access to $\sim \rho(E)$ many different intermediate states in the s-channel. Suppose we estimate the intermediate state with a thermal system with entropy $S\sim \log(\rho)$ then every particular outcome (such as a two particle out-going state) has a probability of $\sim \exp(-\mathcal{O}\left(S)\right)$ which leads to the estimate \eqref{ADOS}. This argument was first made for energies above the black hole threshold in~\cite{Banks:1999gd}.  The idea is that if $E$ is sufficiently large and the impact parameter is sufficiently small, the in-going particles in the $2\to 2$ scattering process are expected to form a black hole with entropy $S_{\rm BH}(E)$. Then, the probability of the black hole to decay into exactly two out-going particles is exponentially suppressed $\left|\mathcal{A}_{2\to 2}(E)\right|^2\sim\exp\left(-\mathcal{O}\left(S_{\rm BH}(E)\right)\right)$.  This has also been confirmed using the Euclidean path integral~\cite{Bah:2022uyz}. 

In addition to the motivation from black hole physics, the estimate \eqref{ADOS} is expected to hold even at energies much below the black hole formation. For example, this identity was verified in string theory~\cite{Bah:2022uyz} to be compatible with the density of massive string states. We take these observations as motivation to estimate $\rho(E)$ for energies above the cutoff via \eqref{ADOS} in general theories of gravity.

Apart from reproducing the expected behavior of $\rho(E)$ in the black hole regime and for perturbative string theory, our estimate in \eqref{ADOS} also addresses the earlier mentioned complications in defining the density of states. Consider first the issue regarding unstable particles. To arrive at our formula for $\rho(E)$, we required it to correctly estimate the density of black hole microstates above the black hole threshold. The black hole microstates, despite being long-lived, are unstable, and they need to be included to give the correct answer. Given that the scattering amplitude includes contributions from unstable particles as well, our estimate for $\rho$ using the amplitude naturally captures the density of states of sufficiently long-lived unstable particles. Note that in this work we are interested in gravitational weak-coupling limits, $\Lambda \ll M_{\rm pl}$, in which the gravitational backreaction of particles below the black hole threshold (with fixed mass in units of $\Lambda$) becomes infinitely small. In this limit, the gravitationally coupled particles become more and more stable and increasingly relevant for any process. As for the issue of weakly coupled bound states, our estimate for $\rho(E)$ captures the spectrum of such bound states as well since they can also be produced as intermediate states in the high-energy scattering process. Finally, regarding the issue of mass averaging, by approximating $\rho(E)$ through scattering amplitudes our estimate in \eqref{ADOS} automatically provides a mass-averaged version of the density of one-particle states.

\subsection{Energy Dependence}\label{ssec:rhoofE}
Given our definition of the one-particle density of states we can now study its dependence on the energy scale $E$. To that end it will prove useful to consider the following ansatz
\begin{equation}\label{rhoofE}
 \rho(E) \sim \exp\left[\left(\frac{E}{\Lambda}\right)^{\alpha}\right] \,.
\end{equation} 
Here $\alpha$ can be viewed as an approximately piece-wise constant function of energy. More precisely we are interested in the scaling of $\log \rho(E)$ up to logarithmic corrections, i.e.
\begin{equation}\label{loglogrho}
    \log \rho(E) = \widetilde{\mathcal{O}}\left[\left(\frac{E}{\Lambda}\right)^{\alpha}\right] 
    \equiv \mathcal{O}\left[\left(\frac{E}{\Lambda}\right)^{\alpha}\cdot \log (E/\Lambda)^k\right]\,,
\end{equation}
where we have used that $\widetilde{\mathcal{O}}$ for a general function $f(x)$ is given by
\begin{align}
    \widetilde{\mathcal{O}}(f(x))=\mathcal{O}(f(x)\cdot\log(x)^k)\,,
\end{align}
for some positive constant $k$. Comparing with~\eqref{ADOS} we see that our approximate definition of $\rho(E)$ in terms of scattering amplitudes is indeed sufficient to capture this leading behavior. When considering particle systems in the following sections it will be easier to first work with  $\Omega(E)$, the density of multiparticle states at energy $E$, instead of $\rho$. Similar to $\rho$ we can make an ansatz 
\begin{equation}
    \log \Omega(E) =\widetilde{\mathcal{O}}\left[\left(\frac{E}{\Lambda}\right)^{\alpha_\Omega}\right]\,. 
    \label{eq:Omega-general}
\end{equation}
Again, $\alpha_\Omega$ is a piece-wise constant function of energy up to logarithmic corrections. Note that, unlike for $\rho$, in order to work with $\Omega$ we need to define an IR-cutoff to make the system finite. This IR-cutoff regulates the infinity coming from the contribution of states obtained by spatially translating a given single-particle state. In this work we consider systems in a box and take this cutoff to be the size of the box.

Importantly, there is a useful connection between $\alpha_\Omega$ and $\alpha$ which allows us to work with $\Omega$ instead of $\rho$ at many points in this work: $\alpha_\Omega\geq1$ implies $\alpha\geq1$. In other words, a sub-exponential growth of $\rho(E)$ cannot lead to an exponential growth of $\Omega(E)$. Before giving a general proof for exponential growths such as \eqref{loglogrho}, it is instructive to study the case where $\rho$ grows polynomially. This case is of particular interest given that KK states in the presence of large extra dimensions have polynomial density. Suppose we have a box of size $L$ with a gas of one-particle states with a density that behaves as $\rho(E) \sim \left(E/E_0\right)^{p-1}$ (e.g. KK modes on a $p$-dimensional compact manifold, with $E_0=m_{\rm KK}$). 
This system has a sub-exponential growth of $\rho$ with energy. Note that the energy dependence of $\rho$ is the same as that of the momentum states of a massless particle in a box with $p$ spatial dimensions of size $1/E_0$. We can use this equivalence to map our computation of multi-particle states to a simpler higher-dimensional problem. The number of multi-particle states in a $(d-1)$-dimensional box of size $L$ arising from a particle spectrum with $\rho(E)\sim \left(E/E_0\right)^{p-1}$ is equivalent to the number of microstates of radiation in a bigger, $(p+d-1)$-dimensional, box consisting of the original $(d-1)$-dimensional box of length $L$ and an additional $p$-dimensional box with sides of length $1/E_0$. Note that in doing so we effectively traded a tower of states with density $\rho(E)$ with $p$ extra compact dimensions that account for that particle density in the $d$-dimensional theory. The entropy of radiation in such a $(p+d-1)$-dimensional box is 
\begin{align}
    S\sim \left(\frac{E}{E_0}\right)^{\frac{p+d-1}{p+d}}\left(E_0L\right)^\frac{d-1}{p+d}\,.
\end{align}
Using $S\sim\log\Omega(E)$ we find
\begin{equation}\label{OmegaKKstates}
    \rho(E) \sim  \left(\frac{E}{E_0}\right)^{p-1} \;\Rightarrow \quad\log \Omega(E) \sim \left(\frac{E}{E_0}\right)^{\frac{p+d-1}{p+d}}\,.
\end{equation}
Thus, a sub-exponential growth of $\rho(E)$ leads to $\alpha_\Omega = (p+d-1)/(p+d)<1$.

The above argument does not yet exclude the possibility of an exponential growth of $\Omega$ due to a one-particle density of states that is superpolynomial but subexponential, i.e. $\rho \sim \exp(E^\gamma)$ for $0<\gamma < 1$. We now give a more general argument that also addresses this case. 

Consider a particle system with total energy $E$ in a large volume $V$ such that the system size is bigger than the size of any particle. Let $\bar{n}$ be the average number of particles in the box. In the thermodynamic limit the actual number of particles in the box for a generic state with energy $E$ will deviate from $\bar{n}$ by 
\begin{equation}\label{deltan}
   \frac{\delta n}{\bar n} \sim \frac{1}{\sqrt{\bar{n}}}\,. 
\end{equation}

We can now provide an upper bound for the entropy of the system at fixed energy $E$. Since in the thermodynamic limit $\delta n \sim \sqrt{\bar{n}}$ the majority of the entropy is accounted for by states with $n\leq \bar{n}+ \sqrt{\bar{n}}$ particles. For definiteness we therefore consider states with $n\leq 2\bar{n}$ particles which will indeed account for the majority of the entropy. On the other hand, the average energy of a particle in the system is $\epsilon=\frac{E}{\bar{n}}$. Again, in the thermodynamic limit the average energy of a particle in the box will deviate from $\epsilon$ only by 
\begin{equation}
    \delta \bar{\epsilon} \sim  \frac{E}{\bar{n}^2}\,\delta{n} \sim  \frac{\epsilon}{\sqrt{\bar n}}\,,
\end{equation}
where we used \eqref{deltan}. An upper bound for the entropy is now given by considering states with $n\leq 2\bar{n}$ particles such that the rest mass $m$ of each particle satisfies $m\leq 2\epsilon$. In addition we allow the $n\leq 2\bar n$ particles to have any momentum inside the box. Since already the rest mass can be of order $2\epsilon$ the total energy of most of these states will exceed $E$. Therefore the number of the states we consider is much larger than the microstates in the microcanonical ensemble at fixed $E$. Ignoring the further overcounting from identical particles, we find the following upper bound for the entropy
\begin{equation}
\begin{aligned}
    S<\log\left[\left(1+\int_{0}^{2\epsilon} dm\, \rho(m)\right)^{2\bar n}(V\Lambda^{d-1})^{2\bar n}\right]
      \lesssim 2\bar n\log\left[\left(1+\int_{0}^{2\epsilon} dm\, \rho(m)\right)\right]+\mathcal{O}(\bar n)\,.
\end{aligned}
\end{equation}
Here $V\Lambda^{d-1}$ accounts for the number of choices for the momentum of a single particle with a UV cutoff $\Lambda$ and IR-cutoff given by the size of the box $V^{1/(d-1)}$. Moreover, $\int_{0}^{2\epsilon} dm\, \rho(m)$ represents the number of particles with rest mass $m\leq 2\epsilon$ of which we choose less than $2\bar n$.\footnote{This can also be viewed as counting the number of lists with $2\bar n $ slots such that in each slot, there is either a particle with mass less than $2\epsilon$, or no particle at all. The no-particle choice for a given slot is accounted for by the additional $1$ in $\left(1+\int_{0}^{2\epsilon} dm\, \rho(m)\right)$.} If $\alpha_\Omega\geq1$, we have $\bar n\epsilon/\Lambda=E/\Lambda\lesssim S$ from which we infer
\begin{align}
    \epsilon/\Lambda\lesssim \log\left[\left(1+\int_{0}^{2\epsilon} dm\, \rho(m)\right)\right]+\mathcal{O}(1)\,.
\end{align}
Therefore, $\left(\int_{0}^{2\epsilon} dm\,\rho(m)\right)$ must have at least an exponential growth in $\epsilon$ which implies that $\rho(m)$ also has at least an exponential growth in its argument. We thus conclude
\begin{align}\label{uin}
    \alpha_\Omega\geq 1\quad \Rightarrow \quad \alpha\geq 1\,.
\end{align}
Note that since the number of multiparticle states $\Omega$ is greater than the number of one-particle states $\rho$, we have $\alpha_\Omega\geq\alpha$. Therefore, using \eqref{uin} we find 
\begin{align}\label{uin2}
    \alpha_\Omega=1\quad\Leftrightarrow\quad\alpha=1
\end{align}
This result turns out to be crucial for our discussion in section~\ref{sec:Hagedorn}. A well-known example of a system that exhibits exponential growth for both, the density of one-particle and the multi-particle states, with energy is given by perturbative strings for which the exponential growth of the string excitations translates into an exponential growth of $\Omega(E)$~\cite{Mitchell:1987hr,Mitchell:1987th}.

\subsection{Connection to particle/black hole transition}\label{sec:particleBH}
Through their relation to the one-particle density of states in \eqref{ADOS} scattering amplitudes contain important information about the high-energy regime of quantum gravitational theories. We now show that in fact they can be used to extract information about the particle/black hole transition. More precisely we show the following 
{\proposition \label{propparticleBH} Consider a $d$-dimensional theory of gravity in asymptotically flat spacetime with quantum gravity cutoff $\Lambda$. The states with mass $E$ are black hole microstates if and only if at energy $E$ the density of one-particle states scales as $\log \rho(E) \sim (E/\Lambda)^\alpha$ with $\alpha>1$.} \\

In order to prove this Proposition we consider Schwarzschild black holes in asymptotically flat $d$-dimensional spacetime. The entropy for these black holes is given by
\begin{equation}\label{entropySchwarzschild}
      S_{\rm BH,d}(E)=\log \rho(E)_{{\rm BH},d} \sim \left(\frac{E}{M_{\rm pl,d}}\right)^\frac{d-2}{d-3}\:, 
\end{equation}
where we have identified $E$ with the mass of the black hole. Comparing~\eqref{entropySchwarzschild} with~\eqref{rhoofE}, we get $\alpha=\frac{d-2}{d-3}>1$ for black holes. This straightforwardly shows that for energies above the black hole threshold $\alpha>1$. Note that in the argument above, we treated the black hole microstates as one-particle states, which fits the definition of $\rho(E)$ in section~\ref{OPS}. 
This is because different microstates of a single black hole can be viewed as different particles with a horizon (e.g. when their Schwarzschild radius is greater than their Compton wavelength). This interpretation is consistent with the computation of the entropy of supersymmetric black holes via counting of BPS states~\cite{Strominger:1996sh}, which shows that the entropy of black holes can microscopically be traced back to the large degeneracy of one-particle states with fixed spin, charge, and mass. 

To show the converse, i.e., $\alpha \leq 1$ implies that states have to be particles, consider the free energy in the canonical ensemble at temperature $T=1/\beta$ in a box of size $\Lambda_{\rm IR}^{-1}$, where $\Lambda_{\rm IR}$ is some IR-cutoff:
\begin{align}\label{eq:freenergy}
    e^{-\beta F}=\int_{\Lambda_{\rm IR}} {\rm d}E\,\Omega(E)\exp(-\beta E)
    \sim \int_{\Lambda_{\rm IR}} {\rm d}E\,\exp(\mathcal{O}(E^{\alpha_\Omega})-\beta E)
    \,.
\end{align}
For any fixed total energy $E$ inside the box the gravitational backreaction vanishes in the gravitational weak-coupling limit and the system is described by particles. In other words, the particle description emerging in such limits must make sense for arbitrarily high total energies which implies that the unbounded partition function must converge. However, the convergence of the integral in~\eqref{eq:freenergy} requires $\alpha_\Omega \leq 1$. As we explained in section~\ref{ssec:rhoofE}, $\alpha_\Omega \leq 1$ implies $\alpha \leq 1$ which concludes our proof of Proposition~\ref{propparticleBH}. 

Using this result the particle/black hole transition can be read off from the scattering amplitudes in the following way. The scattering amplitudes are related to the density of one-particle states via \eqref{ADOS}. Therefore if we track the scattering amplitude as a function of energy we obtain an approximate form of $\rho(E)$ as in \eqref{rhoofE}. As a consequence of Proposition~\ref{propparticleBH} the energy at which the particle/black hole transition takes place corresponds to the energy at which $\alpha$ in \eqref{rhoofE} switches from $\alpha\leq 1$ to $\alpha>1$. Therefore knowledge of the scattering amplitudes as a function of $E$ allows us to read off the energy scale $M_{\rm min}$ at which the particle/black hole transition takes place. The energy $M_{\rm min}$ corresponds to the minimal mass of any black hole in the theory. Importantly, as we discuss in more detail in section~\ref{sec:higherdimBH}, in the presence of compact extra dimensions this minimal mass black hole is in fact a higher-dimensional black hole.

\section{Species Scale and Gravitational Amplitudes}\label{spscandamplitudes}

In this section we discuss the species scale, i.e. the scale suppressing the higher-derivative corrections to the gravitational effective action, from the perspective of gravitational amplitudes.  In section~\ref{sec:ExistenceEFT} we first argue that the scale appearing in the higher-derivative corrections is indeed the cutoff of the gravitational EFT defined as the inverse radius of the smallest black hole in the theory. We do this by matching the $2\to2$ scattering amplitude of gravitons in a certain regime with the expansion of the effective action. In section~\ref{sec:boundtension} we further use the contribution to gravitational amplitudes from effective strings to constrain the tension of weakly coupled $p$-branes with $p\geq 1$ in theories of quantum gravity in terms of the species scale. 

Before we discuss the above points in detail, let us pause and clarify what we mean by effective field theory in the following. To define an effective field theory one typically fixes a spacetime dimension, the field content of the theory and the symmetries of the theory. In addition one specifies a cutoff $\Lambda_{\rm EFT}$ determining the regime of validity of the EFT description. For example we can consider a $d$-dimensional theory of Einstein gravity coupled to a set of light fields exhibiting certain symmetries. This theory has a cutoff $\Lambda_{\rm EFT}$ beyond which the field theory description breaks down. This breakdown can, for example, be due to the appearance of a tower of states that cannot consistently be described in terms of an EFT. The classical example for such a situation is a $d$-dimensional theory that arises from a higher-dimensional theory. For energies above the KK scale, KK modes need to be included in the EFT. This is however not possible in a parametrically controlled way, since e.g. for scattering process with two initial particles of mass $\Lambda_\text{KK} = 1/R$, the amplitude becomes non-unitary and one must include the second KK particle of mass $2/R$, even at zero initial momentum.\footnote{Further, even if we were to include a finite number of KK modes, they become strongly coupled at the scale $\Lambda_\text{QG}$, so that the EFT cutoff cannot be made higher than $\Lambda_\text{QG}$.} However, this does not mean that there does not exist any EFT description at energies above $\Lambda_{\rm KK}$ since we can describe the theory in terms of the higher-dimensional EFT. Of course the field content and the symmetries of this EFT are different from the original $d$-dimensional theory but this new EFT allows us to describe the infinitely many KK-states in terms of finitely many (higher-dimensional) fields. The cutoff of this new theory is expected to be above $\Lambda_{\rm KK}$ but the quantum gravity cutoff $\Lambda$, defined in terms of the higher-derivative corrections, does not change under dimensional reduction. Hence even though the original $d$-dimensional gravitational EFT is not valid above $\Lambda_{\rm KK}$ there may exist \emph{some} EFT description that is valid up to $\Lambda$ which -- in the case of a KK-reduction -- is necessarily higher-dimensional. It is in this sense that in this section we show that, indeed, there always exists some EFT description with cutoff $\Lambda_{\rm EFT}=\Lambda_{\rm QG}$ even though this EFT may significantly differ from the original $d$-dimensional EFT.

\subsection{Existence of a (higher-dimensional) EFT with cutoff \texorpdfstring{$\Lambda$}{}}\label{sec:ExistenceEFT}
Consider a theory of gravity at some energy $E>M_{\rm min}$ where $M_{\rm min}$ is the energy scale at which the scaling of the density of one-particle states in \eqref{rhoofE} changes from $\alpha\leq 1$ to $\alpha>1$. From Proposition~\ref{propparticleBH} we recall that for energies $E>M_{\rm min}$ the one-particle states are gravitationally strongly coupled states whose mass is dominated by their gravitational self-energy. In other words these states correspond to black holes. In Lorentzian signature such states are hence expected to be describable as solutions of a Lorentzian gravitational field theory with a horizon. In order for the gravitational field theory description to be applicable, the curvature of the horizon furthermore needs to be smaller than the cutoff of the field theory.
Consequentially, there needs to exist a field theory with cutoff $\Lambda_{\rm min}$ such that all states with mass $M>M_{\rm min}$ are describable as field theory solutions for which the curvature at the horizon does not exceed $\Lambda_{\rm min}^2$. This implies that the horizon radius of the smallest black hole in the theory should not be smaller than $\Lambda_{\rm min}^{-1}$. As mentioned at the beginning of this section let us stress again that if we start with a $d$-dimensional EFT, the actual EFT with cutoff $\Lambda_{\min}$ can be higher-dimensional. In this case the black hole with minimal mass $M_{\rm min}$ is a black hole of this higher-dimensional EFT and is localized in the extra dimension.

On the other hand, as reviewed in section~\ref{ssec:speciesscale}, the quantum gravity cutoff, or the species scale, $\Lambda$, is defined as the scale that suppresses the higher-derivative terms in the effective gravitational action \eqref{scorr}. Based on EFT arguments, one expects the higher derivative corrections at cut-off scale $\Lambda_{\min}$ to be controlled by the same energy scale ($\Lambda\sim\Lambda_{\min}$).  In the following, we give an independent argument based on scattering amplitudes and the properties of black hole physics that makes this connection more clear. 
{\proposition \label{prop:existenceofEFT} Consider a theory of quantum gravity in which the smallest black hole (which could be higher-dimensional) has radius $\Lambda_{\rm min}^{-1}$ and mass $M_{\rm min}$. Then the higher-derivative corrections to the Einstein--Hilbert action are of the form 
\begin{equation}
\mathcal{L}_{\rm corr} \supset \frac{M_{\rm pl,d}^{d-2}}{\Lambda_{\rm min}^{2n}}\mathcal{O}_{2n+2}(\mathcal{R})\,,
\end{equation}
where $\mathcal{O}_m(\mathcal{R})$ is a dimension-$m$ operator involving $\mathcal{R}$ and its derivatives.}\\

Let us make an important clarifying remark. As pointed out in~\cite{Bedroya:2024uva}, the species scale $\Lambda$, and the inverse radius of the smallest black hole described by a given EFT can be vastly different. However, here we allow for black hole solutions to any effective action that is valid at low energies. In particular this includes higher-dimensional black holes that are localized in possible extra dimensions.

Given the definition of the quantum gravity cutoff $\Lambda$ in terms of the higher-derivative corrections in section~\ref{ssec:speciesscale}, the main takeaway of Proposition~\ref{prop:existenceofEFT} is that there indeed exists some effective field theory description up to energies of order of the quantum gravity cutoff $\Lambda$. Let us stress that the statement of Proposition~\ref{prop:existenceofEFT} has previously appeared in the literature. However, as we will discuss in the following, there are some subtleties in the justification of the identification between $\Lambda$ and $\Lambda_{\rm min}$. Typically one argues in the following way \cite{vandeHeisteeg:2023dlw,vandeHeisteeg:2023ubh}: suppose there is a perturbative series of higher-derivative corrections, each proportional to $M_{\rm pl,d}^{d-2}\Lambda^{-2n}\mathcal{R}\Box^{n-1}\mathcal{R}$ or $M_{\rm pl,d}^{d-2}\Lambda^{-2n}\mathcal{R}^{n+1}$, then the effective action containing this series diverges for $\mathcal{R}\gg \Lambda^2$. Since $\Lambda_{\rm min}^2$ is the curvature scale at the horizon of the smallest black hole that can be described in effective field theory, one expects the breakdown of the higher-derivative expansion to occur for curvatures of the order of $\Lambda_{\rm min}^2$ which would imply $\Lambda=\Lambda_{\rm min}$. 
This reasoning is, however, flawed for two reasons:
\begin{enumerate}
\item The same argument could be applied to non-gravitational field theories with a series of irrelevant operators to argue for an ultimate cutoff for the EFT description. However, that conclusion would be incorrect. For example, the effective action of quantum electrodynamics at energies below the electron mass $m_e$ has irrelevant operators $\propto (m_e)^{-4n+d}(F^2)^n$. Nevertheless in this case the breakdown of the EFT at energies $\sim m_e$ is attributed to the fact that additional massive modes which were previously ignored (in this case the electron) need to be integrated in. In this way, in field theory the cutoff can be raised by integrating in additional states. The same is, however, not possible in theories of gravity where, e.g., black hole microstates are expected to be strongly coupled and cannot be simply integrated into the EFT. This implies that the quantum gravity cutoff is qualitatively different from some field theory cutoffs. 
\item The effective action does not necessarily diverge at curvature scales $\mathcal{R}\gg\Lambda^{2}$ as one might have expected. For example, consider the higher-derivative term
\begin{equation} \label{Lcorr}
\mathcal{L}_{\rm corr} \supset \hat{a}_n \frac{M_{\rm pl,d}^{d-2}}{\Lambda^{2n}}\mathcal{R}\Box^{n-1}\mathcal{R}\,,
\end{equation}
where we have included a constant $\hat{a}_n$ here that accounts for a possible $n$-dependence of the Wilson coefficient of the above term and does not include any dependence on the scalar fields in the theory. Near the horizon of a black hole with radius $r_H$, we have $\mathcal{R}\Box^{n-1}\mathcal{R}\propto r_H^{-2n-2}$. One may expect that the sum of such terms diverges for $r_H\ll\Lambda^{-1}$. However, as we will shortly see, the coefficients $\hat a_n$ are exponentially small in $n$ which can make the series convergent. This is due to the fact that gravitational $2\rightarrow2$ scattering amplitudes are always exponentially suppressed at high energies and the higher-derivative terms are just a Taylor expansion of such amplitudes in energy, see Appendix~\ref{app:amplitude}. The higher-derivative expansion of the gravitational effective action hence does not necessarily diverge at curvature scales $\mathcal{O}(\Lambda^2)$.
\end{enumerate}

Having explained the shortcomings of the usual argument we need to work a bit harder in order to establish $\Lambda=\Lambda_{\rm min}$ and argue for Proposition~\ref{prop:existenceofEFT} correctly. To that end we make use of certain properties of gravitational amplitudes which we summarize in Appendix~\ref{app:amplitude}. In short, we show that the imprint of the smallest black hole on high-energy $2\rightarrow 2$ gravitational amplitude can be reproduced only by higher-derivative corrections of the form
\begin{align}\label{LcorrLambdaMin}
    \mathcal{L}_{\rm corr} \supset \hat{a}_n \frac{M_{\rm pl,d}^{d-2}}{\Lambda_{\min}^{2n}} \mathcal{R}\Box^{n-1}\mathcal{R}\,,
\end{align}
which, by comparison to \eqref{Lcorr} implies $\Lambda_{\rm min}=\Lambda$.

To argue for \eqref{LcorrLambdaMin} we consider $2\to 2$ scattering amplitudes for gravitons. At center of mass energies $E\gg M_\text{min}$ and impact parameter $b \ll r_\text{min}=1/\Lambda_{\rm min}$ the scattering process involves a black hole formation/evaporation process. The exclusive $2\rightarrow 2$ amplitude corresponds to the case where the Hawking radiation consists of only two particles. Therefore, the amplitude is inversely proportional to the number of black hole microstates and is exponentially suppressed in the black hole entropy. Instead of the impact parameter $b$ we are interested in the dependence of the amplitude in terms of the Mandelstam variables $s,t$ defined in \eqref{mandelstam}. To that end we need to perform a Fourier transform on the impact parameter $b$. At fixed angle and large energies, the contribution of black holes to the scattering amplitudes is exponentially suppressed and the dominant contribution comes from particles whose separation is much larger than the Schwarzschild radius, $r_s(E)$, of a black hole with mass $E$. Nonetheless, as explained in appendix \ref{app:unphysical}, the exponential suppression at low-impact parameters still leaves a phase factor in the amplitude which becomes exponentially large upon analytic continuation to the unphysical regime $t\gg 0$ (cf. the discussion around \eqref{UPA})
\begin{equation}\label{bha}
    \mathcal{A} \sim \exp\left(2r_s(E) \sqrt{t}\right)\,,\qquad \text{for} \quad t\gg 0\,. 
\end{equation}
As we detail in appendix~\ref{app:unphysical}, we expect the above behavior to persist at energies $E \ll M_{\min}$. However, $r_s$ will be replaced by $b_c(E)=\Lambda_{\min}^{-1}\mathcal{O}(E/\Lambda_{\min})$.
\begin{align}\label{expamplitude}
        \mathcal{A}\sim \exp\left( \Lambda_{\rm min}^{-1}\sqrt{t} \times \mathcal{O}(\log E/\Lambda_{\rm min})\right)\,,
\end{align}
To reach this conclusion we use that Regge behavior dictates that the energy dependence of $b_c(E)$ at energies $E\ll M_{\rm min}$ can be at most logarithmic and that the amplitude needs to correctly match the black hole result \eqref{bha} at the black hole threshold $E\sim M_{\min}$. 

Since black holes are saddles of the Euclidean gravitational action they can be thought of as non-perturbative contributions to gravitational amplitudes. Nevertheless at energies below $\Lambda_{\rm min}$ the amplitude has to be computable perturbatively from the full effective action including the higher-derivative corrections. The amplitude computed from the full effective action has to match the high-energy amplitude at $\sqrt{s}\sim\mathcal{O}(\Lambda_{\rm min})$ even in the unphysical regime for $t$. To reproduce the exponential amplitudes of the form \eqref{expamplitude} we need an infinite series of higher-derivative terms. For the particular case of the $2\to 2$ scattering amplitude, we hence require terms in the effective action of the form 
\begin{equation}\label{higherder2to2}
    \cL_{\rm eff} \supset M_{\rm pl}^{d-2}\frac{\hat{a}_n}{\Lambda_{\rm min}^{2n}}\mathcal{R}\Box^{n-1}\mathcal{R}\,,
\end{equation} 
 in order to reproduce the amplitude in \eqref{expamplitude}. As before, the $\hat{a}_n$ encode the $n$-dependence of the coefficients of the higher-derivative terms. In order to make sure that the high-derivative terms correctly reproduce the Taylor expansion of \eqref{expamplitude} we need to have 
 \begin{equation}\label{hatan}
 \hat{a}_n\sim \frac{1}{(2n)!(n-1)!}\,.
 \end{equation}
 The $2n!$ comes from the Taylor expansion of the exponential in \eqref{expamplitude} and the $(n-1)!$ is to counter the repetition resulted from having $n-1$ identical d'Alembertians in \eqref{higherder2to2}. From \eqref{higherder2to2} we conclude that the higher-derivative terms are controlled by $\Lambda_{\rm min}$. However, as reviewed in section~\ref{ssec:speciesscale}, the energy scale controlling the higher-derivative expansions was defined to be the species scale or quantum gravity cutoff $\Lambda$ \cite{vandeHeisteeg:2022btw}. Therefore, using amplitudes we showed that $\Lambda=\Lambda_{\rm min}$ and hence Proposition~\ref{prop:existenceofEFT}. Note that compared to the definition~\eqref{deflambda}, here we fixed the overall $n$-dependence of the Wilson coefficients $a_n(\phi)$ such that the scale appearing in the individual higher-derivative terms in \eqref{higherder2to2} is indeed always the species scale.

Let us further note that, as advertised above, contrary to the field theory expectation the sum over the terms in \eqref{higherder2to2} does not diverge at curvature scales $\mathcal{R}\sim \Lambda^{2}$ since the $n$-dependence of the coefficients $\hat{a}_n$ ensures that the radius of convergence of the higher-derivative expansion in \eqref{higherder2to2} is larger than $\mathcal{R}\sim \Lambda^2$. In the context of string theory the suppression in $n$ for instance ensures that string amplitudes are analytic at high energies. 

\subsection{Bound on the tension of weakly coupled \texorpdfstring{$p$}{}-branes \texorpdfstring{$\cT\gtrsim \Lambda^{p+1}$}{}}\label{sec:boundtension}

In section \ref{ssec:speciesscale} we reviewed the definition of the species scale $\Lambda$ in terms of the Wilson coefficients of the higher-derivative corrections to the Einstein--Hilbert action. In this section, we use this definition for $\Lambda$ to bound the tension of all weakly coupled $p$-branes with $p\geq 1$ as 
{\proposition \label{prop:tension} In an effective theory of gravity, the tension $\cT$ of any weakly coupled $p$-brane with $p\geq 1$ is bounded by the quantum gravity cutoff $\Lambda$ as 
\begin{equation}\label{tensionbound}
    \cT \gtrsim \Lambda^{p+1}\,. 
\end{equation}}

In the following we always implicitly assume $p\geq 1$. In this work, we adapt the following definition for a weakly coupled brane: A weakly coupled brane has a self-energy that is negligible compared to its tension. For example, for a $p$-brane  coupled to a $(p+1)$-form gauge potential, this condition implies that the gauge coupling in the units of the tension of the brane is negligible. To be more precise, let us consider such a $p$-brane coupled to a $(p+1)$-form gauge potential $\cA_{p+1}$ for which the kinetic term in the  $d$-dimensional bulk action reads 
\begin{equation}
    S_{p} = -\frac{1}{2g_{p+1}^2}\int \cF_{p+2} \wedge * \cF_{p+2}\,,
\end{equation}
where $\cF_{p+2}$ is the field strength of $\cA_{p+1}$. The gauge coupling $g_{p+1}$ has mass dimension $p+2-d/2$. A typical configuration of the gauge potential on the $p$-brane is of the form 
\begin{equation}
    \cA_{p+1}\sim \frac{g_{p+1}^2}{L^{d-p-3}}\,,
\end{equation}
where $L$ is the typical length scale of the $p$-brane. The self-energy associated to the coupling of the $p$-brane to $\cA_{p+1}$ can now be estimated as 
\begin{equation}
    \delta E \propto \cA_{p+1} L^p \sim \frac{g_{p+1}^2}{L^{d-2p-3}}\,. 
\end{equation}
The requirement that the self-energy induced by the coupling of the $p$-brane to $\cA_{p+1}$ is negligible compared to the tension then amounts to 
\begin{equation}\label{tensioncouplingweak}
    \delta E < \cT L^p \qquad \Rightarrow \qquad \frac{g_{p+1}^2}{L^{d-p-3}}<\cT\,. 
\end{equation}
Given our definition of weakly coupled $p$-branes we start by considering the case of weakly coupled strings with tension $\cT$. It is instructive to review why the inequality $\cT\gtrsim \Lambda^2$ holds in the string landscape. The effective action of any perturbative string theory, at tree-level, contains a series of higher-derivative corrections 
\begin{align}\label{SEA}
    S_{\rm tree} = \frac{M_{\rm pl,d}^{d-2}}{2} \int {\rm d}^d x \sqrt{-g} \left(\mathcal{R}+\sum_{n=4}^{\infty} \hat{a}_n\frac{\mathcal{O}_n(
\mathcal{R})}{\cT^{\frac{n-2}{2}}}\right)+\hdots\,,
\end{align}
where $\hat{a}_n$ are dimensionless constants that do not depend on any scalar fields.\footnote{Any dependence of the Wilson coefficients of the higher-derivative terms on the scalar fields in the theory will be encoded in $\mathcal{T}$ directly.} This expansion is known as the $\alpha'$-expansion in string theory. By matching the above action with the expression \eqref{Lcorr}, we can identify the species scale as $\Lambda\sim \sqrt{\cT}$ in any perturbative string theory. Note that the effective action in string theory also contains a sub-leading series of loop corrections which are suppressed by positive powers of the string coupling $g_s$. Therefore, in the weak-coupling limit, $g_s\ll1$, the effect of the loop expansion on $\Lambda$ is negligible. 

Let us revisit how the tree-level string amplitude gives rise to the perturbative series in the effective action \eqref{SEA}. We first identify the imprint of higher-derivative corrections on high-energy $2\rightarrow 2$ amplitudes from which we read off the corresponding Wilson coefficients. Consider for example a term of the form $\frac{M_{\rm pl,d}^{d-2}}{M^{2k}}\mathcal{R}\Box^{k-1}\mathcal{R}$ in the effective action where $M$ is some mass scale. We expand the spacetime metric $g_{\mu\nu}$ around Minkowski background $\eta_{\mu\nu}$ as 
\begin{align}\label{gmunuhmunu}
    g_{\mu\nu}=\eta_{\mu\nu}+\frac{h_{\mu\nu}}{M_{\rm pl,d}^\frac{d-2}{2}},
\end{align}
where the normalization of $h$ ensures that it has a canonical kinetic term. The higher-derivative term gives rise to  a four-point interactions of the form
\begin{equation}
   \mathcal{R}\Box^{k-1}\mathcal{R} \;\longrightarrow \; M_{\rm pl,d}^{4-2d}\partial_\gamma h_{\alpha \beta}\partial^{\beta}h^{\alpha\gamma}\Box^{k-1}\partial_{\gamma'} h_{\alpha' \beta'}\partial^{\beta'}h^{\alpha'\gamma'}
\end{equation}
which has $2k+2$ derivatives and for which the coefficient in the effective action is proportional to $M_{\rm pl,d}^{d-2}/M^{2k}$. Therefore, such a term contributes to the fixed-angle $2\rightarrow 2$ graviton amplitude in the following way
\begin{align}\label{PLT}
    \frac{M_{\rm pl,d}^{d-2}}{M^{2k}}\mathcal{R}\Box^{k-1}\mathcal{R}\in\mathcal{L}~~~\rightarrow~~~\frac{E^{2k+2}}{M^{2k}M_{\rm pl,d}^{d-2}}f(\theta)\in \mathcal{A}(E,\theta),
\end{align}
where $E$ is the center of mass energy and $\theta$ is the scattering angle. We have neglected the polarization-dependence  for the sake of simplicity. Knowing how each term contributes to the gravitational amplitudes, allows us to read off the Wilson coefficient $M_{\rm pl,d}^{d-2}/M^{2k}$ from the amplitude in string theory. In string theory, the $2\rightarrow 2$ amplitude at tree-level is given by the high-energy limit of the Virasoro-Shapiro amplitude \cite{Polchinski:1998rq}
\begin{equation}\label{sphere}
    \mathcal{A}(E,\theta)\sim \frac{\cT}{M_{\rm pl,d}^{d-2}}\,\exp\left[-\frac{1}{16\pi}\left(\frac{s/\cT}{\log(s/\cT)}+\frac{t/\cT}{\log(t/\cT)}+\frac{u/\cT}{\log(u/\cT)}\right)\right].
\end{equation}
where $(s,t,u)$ are the Mandelstam variables and $\cT$ is the string tension. Note that, if we go to energies much higher than $\sqrt{\cT}$ the effect of higher-genus contributions will become dominant and change the exponential behavior of the amplitude from $\exp(cE^2/\log(E))$ to $\exp(-\mathcal{O}(E))$ (see appendix \ref{GMSA}). However, our conclusion about the perturbative expansion of the amplitude in terms of $E/\sqrt{\cT}$ will not be affected by the higher-genus corrections. The contribution in \eqref{sphere} comes from the sphere amplitude which can also be understood as a saddle of the Polyakov action ($\mathcal{A}\propto\exp(S_\text{saddle})$) \cite{Gross:1987ar}. If one expands the above amplitude in terms of the contribution in \eqref{PLT} of local terms to the full amplitude one finds
\begin{equation}
    \mathcal{A}(E,\theta)= \sum_{k\geq1} f_k(\theta)\frac{E^{2k+2}}{\cT^{k}M_{\rm pl,d}^{d-2}}\,,
\end{equation}
where we absorbed the $k$-dependend expansion coefficients into $f_k(\theta)$. Therefore, the Wilson coefficient of $\mathcal{R}\Box^{k-1}\mathcal{R}$ in any perturbative string theory is proportional to $(M_{\rm pl,d}^{d-2}/\cT^{k})$. The above argument shows how the sphere amplitude ensures the presence of higher-order corrections of the type $\frac{M_{\rm pl,d}^{d-2}}{\cT^{k}}\mathcal{R}\Box^{k-1}\mathcal{R}$ consistent with \eqref{SEA}. Therefore in critical string theory the species scale is indeed set by $\Lambda\sim \sqrt{\cT}$.

Having reviewed how the tension of a critical string is related to the species scale, let us revisit the above argument and understand to what extent the argument depends on the criticality of the string. There are three assumptions underlying the argument which are met by the critical string but not necessarily true for a generic string:
\begin{enumerate}
    \item The graviton is a string state and therefore the gravitational amplitude is given by a string amplitude. 
    \item The action of the string is known exactly and does not receive corrections based on the embedding of the string worldsheet in the spacetime.
    \item The tree-level amplitude (i.e. sphere diagram) is the dominant contribution. 
\end{enumerate}

In the following, we argue that one can reproduce the same argument for any \emph{weakly coupled} string. More precisely, we show that any weakly coupled string with tension $\cT$ will generate higher-derivative corrections 
\begin{equation}\label{eq:Scorrweakstring}
    S_{\rm corr} \propto \int d^dx \sqrt{g}  \frac{M_{\rm pl,d}^{d-2}}{\cT^{k}}\mathcal{R}\Box^{k-1}\mathcal{R}\,.    
\end{equation} 
As reviewed in section~\ref{ssec:speciesscale}, the overall magnitude of the Wilson coefficients of the higher-derivative terms sets the species scale. Therefore the term in \eqref{eq:Scorrweakstring} has to be at least suppressed by the species scale leading to the bound
\begin{align}\label{eq:Tgtrlambda}
    \frac{1}{\Lambda^{2k}}\gtrsim\frac{1}{\cT^k}\;\Rightarrow \; \cT\gtrsim \Lambda^2\,,
\end{align}
which is \eqref{tensionbound} for the case of strings, i.e. $p=1$. Let us stress that this bound is consistent with the analysis in~\cite{Cota:2022yjw} where it was found that in F-theory compactification on elliptically fibered Calabi--Yau fourfolds the tension of any weakly coupled string, dubbed EFT strings in \cite{Lanza:2021udy}, is bounded from below by the species scale in asymptotic limits; see also \cite{Martucci:2024trp} for a discussion of the relation between species scale and EFT string tensions in such setups.

Let us start with the definition of a weakly coupled string (we refer the reader to appendix~\ref{A1} for some details on general effective strings). A weakly coupled string is a string that has a Nambu--Goto action with corrections that are suppressed as long as the extrinsic curvatures $L^{-2}$ of the worldsheet are smaller than $\cT$. In other words the coupling of a weakly coupled string to itself via any gauge field (e.g. graviton or a two-form) can be ignored as long as the string is in a sufficiently flat, or long wave-length, configuration.\footnote{For example for the QCD string, this assumption is violated close to the confinement scale where the string self-intersects and becomes highly curved. Moreover, the effective action of the QCD string is expected to depend strongly on its extrinsic curvature \cite{Gross:1992tu}.} In such a limit, the physics of the string is described by the IR limit of the worldsheet theory. Therefore a weakly coupled string is a string that, in configurations with curvature smaller than its tension, is described in the IR by a two-dimensional CFT. Since we expect the string action to contain a Nambu--Goto term, it is crucial to understand the two-dimensional CFT corresponding to the Nambu--Goto action for non-critical strings.

For the pure Nambu--Goto action this CFT was worked out in \cite{Polchinski:1991ax}, by adding a term to the Polyakov action that cancels the conformal anomaly. In the light-cone gauge the action is given by
\begin{align*}
    S=\int {\rm d}\sigma^+{\rm d}\sigma^{-}\left[\cT\partial_+X^{\mu}\partial_-X_\mu+\frac{d-26}{12}\frac{(\partial_+^2X^\mu\partial_-X_\mu)(\partial_-^2X^\nu\partial_+X_\nu)}{(\partial_+X^\rho\partial_-X_\rho)^2}+\mathcal{O}(L^{2}/\cT)\right]\,.
\end{align*}
For a mildly curved string ($L\sim \partial_{\pm}X^{\mu}\ll \cT^{-1/2}$) the second term is also negligible. We are interested in the contribution of the weakly coupled effective strings to the gravitational scattering amplitudes. For the fundamental strings in string theory, the high energy behavior can be deduced from the worldsheet saddle with the dominant contribution. 

In the following, we apply the same analysis to the effective string in the regime of its validity.\footnote{Our considerations do not apply to the weakly coupled string description of large N QCD, since the strings are not fundamental and are resolved into the constituent field theory states above the confinement scale.} For the effective description to be valid, we focus on worldsheet saddles with weak extrinsic curvatures $L\sim \partial_{\pm}X^{\mu}\ll \cT^{-1/2}$ for which the Polyakov action is a good classical approximation. Some saddles of the Polyakov action, which are $N$-covering of a sphere with punctures, were studied by Gross and Mende \cite{Gross:1987ar}, see appendix~\ref{GMSA} for a review. The tree-level saddle is given by the following embedding
\begin{equation}
    X^\mu=\sum_{j=1}^4\frac{ip_j^\mu}{4\pi N \cT}\log|z-z_j|+\mathcal{O}\left(\frac{\sqrt{\cT}}{s}\right)\,,
\end{equation}
where $p_j$ are the momenta of ingoing/outgoing particles and $z$ is a complex coordinate parametrizing a complex sphere with a point removed and for which the worldsheet metric has been Weyl rescaled to be flat. Moreover, the locations $z_j$ of the external legs on the sphere satisfy the following relation in the saddle configuration
\begin{align}
    \frac{(z_1-z_3)(z_2-z_4)}{(z_1-z_2)(z_3-z_4)}\simeq \frac{-t}{s}\,.
\end{align}
As explained in appendix~\ref{GMSA}, there is a range of $N$ for which we have
\begin{equation}\label{SAM}
    L\gg \frac{1}{\sqrt{\cT}}~~\text{and}~~S_{\rm Saddle}\sim \frac{E}{\sqrt{\cT}}\widetilde{\mathcal{O}}\left(L\sqrt{\cT}\right)\,,
\end{equation}
where $S_{\rm Saddle}$ is the action of the saddle. For these saddles, the external legs of the worldsheet are string states (see figure \ref{GMS}). However, since we do not assume that the graviton is a string state, we need to cut the saddle and add a graviton vertex according to its interaction through the Nambu--Goto term 
\begin{equation}
\int {\rm d}\sigma^+{\rm d}\sigma^{-} \frac{\cT}{M_{\rm pl,d}^\frac{d-2}{2}}h_{\mu\nu}\partial_+X^{\mu}\partial_-X^\nu\,,
\end{equation} 
where we used the normalization in \eqref{gmunuhmunu}. The above term gives an interaction vertex 
\begin{equation}\label{VFD}
    V_{\rm graviton-worldsheet}\sim\frac{\cT L^2}{M_{\rm pl,d}^\frac{d-2}{2}}\,.
\end{equation}
Since we need to insert four such vertices, we find that the contribution of the worldsheet saddle to the gravitational amplitude is 
\begin{align}
    \mathcal{A}_{\rm string}\sim  V_{\rm graviton-worldsheet}^4\cT^\frac{d}{2}e^{-S_{\rm saddle}}\,,
\end{align}
where the factor $\cT^\frac{d}{2}$ comes from the integration measure of the worldsheet action.\footnote{For example, if we were to calculate the contribution of the worldsheet saddle to the cosmological constant, it would go like $\cT^\frac{d}{2}e^{-S_{\rm saddle}}$.} Plugging the saddle action~\eqref{SAM} and the vertex amplitude~\eqref{VFD} into the above expression for $\cA_{\rm string}$ leads to
\begin{align}
    \mathcal{A}_{\rm string}\sim \frac{\cT^\frac{d+8}{2} L^8}{M_{\rm pl,d}^{2(d-2)}}\exp\left[-\frac{E}{\sqrt{\cT}}\widetilde{\mathcal{O}}\left(L\sqrt{\cT}\right)\right]\,. 
\end{align}
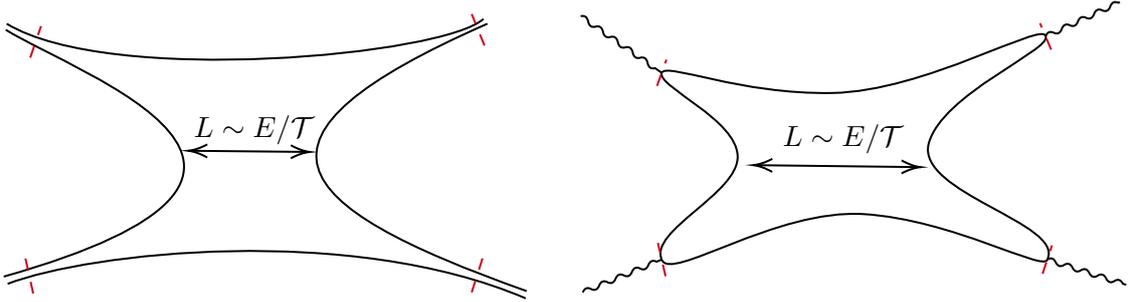
\begin{figure}
    \centering

\tikzset{every picture/.style={line width=0.75pt}} %set default line width to 0.75pt        

\begin{tikzpicture}[x=0.75pt,y=0.75pt,yscale=-1,xscale=0.87]
%uncomment if require: \path (0,210); %set diagram left start at 0, and has height of 210

%Curve Lines [id:da6788251391993942] 
\draw    (9.69,33.87) .. controls (69.35,65.51) and (260.72,49.69) .. (283.23,31.43) ;
%Curve Lines [id:da5011594391621736] 
\draw    (9.13,36.91) .. controls (51.9,60.03) and (216.82,109.32) .. (8,161.05) ;
%Curve Lines [id:da9361781418413941] 
\draw    (10.25,164.7) .. controls (57.53,148.88) and (206.69,134.27) .. (307.44,172) ;
%Curve Lines [id:da7579794510824676] 
\draw    (285.49,33.87) .. controls (136.89,91.07) and (167.29,124.54) .. (308,168.35) ;
%Straight Lines [id:da29484415896978877] 
\draw    (115.82,97.78) -- (181.61,98.35) ;
\draw [shift={(183.61,98.37)}, rotate = 180.5] [color={rgb, 255:red, 0; green, 0; blue, 0 }  ][line width=0.75]    (10.93,-3.29) .. controls (6.95,-1.4) and (3.31,-0.3) .. (0,0) .. controls (3.31,0.3) and (6.95,1.4) .. (10.93,3.29)   ;
\draw [shift={(113.82,97.76)}, rotate = 0.5] [color={rgb, 255:red, 0; green, 0; blue, 0 }  ][line width=0.75]    (10.93,-3.29) .. controls (6.95,-1.4) and (3.31,-0.3) .. (0,0) .. controls (3.31,0.3) and (6.95,1.4) .. (10.93,3.29)   ;
%Straight Lines [id:da0011742456843391036] 
\draw [color={rgb, 255:red, 208; green, 2; blue, 27 }  ,draw opacity=1 ] [dash pattern={on 4.5pt off 4.5pt}]  (23.2,50.91) -- (29.39,35.09) ;
%Straight Lines [id:da6199327966030515] 
\draw [color={rgb, 255:red, 208; green, 2; blue, 27 }  ,draw opacity=1 ] [dash pattern={on 4.5pt off 4.5pt}]  (24.89,169.57) -- (19.82,149.49) ;
%Straight Lines [id:da3173056276192292] 
\draw [color={rgb, 255:red, 208; green, 2; blue, 27 }  ,draw opacity=1 ] [dash pattern={on 4.5pt off 4.5pt}]  (276.48,168.96) -- (282.67,151.92) ;
%Straight Lines [id:da2513007544919663] 
\draw [color={rgb, 255:red, 208; green, 2; blue, 27 }  ,draw opacity=1 ] [dash pattern={on 4.5pt off 4.5pt}]  (283.8,44.82) -- (276.48,29) ;
%Curve Lines [id:da8311922321678133] 
\draw    (385.5,58.55) .. controls (388.44,53.21) and (429.95,69.53) .. (483.39,68.7) .. controls (536.82,67.88) and (604.77,32.14) .. (605.75,40.27) ;
%Curve Lines [id:da08417211285335746] 
\draw    (385.5,58.55) .. controls (382.66,65.57) and (428.85,81.97) .. (429.06,100.7) .. controls (429.27,119.42) and (378.16,137.77) .. (385.5,152.5) ;
%Curve Lines [id:da858777416226024] 
\draw    (385.5,152.5) .. controls (390.39,164.18) and (450.66,126.85) .. (501.5,129.64) .. controls (552.33,132.43) and (604.7,158.81) .. (606.73,152.5) ;
%Curve Lines [id:da08743904168405803] 
\draw    (605.75,40.27) .. controls (609.15,47.05) and (546.24,69.23) .. (538.69,94.1) .. controls (531.15,118.96) and (617,134.21) .. (606.73,152.5) ;
%Straight Lines [id:da4051007824281785] 
\draw    (440,105.02) -- (532,105.98) ;
\draw [shift={(534,106)}, rotate = 180.6] [color={rgb, 255:red, 0; green, 0; blue, 0 }  ][line width=0.75]    (10.93,-3.29) .. controls (6.95,-1.4) and (3.31,-0.3) .. (0,0) .. controls (3.31,0.3) and (6.95,1.4) .. (10.93,3.29)   ;
\draw [shift={(438,105)}, rotate = 0.6] [color={rgb, 255:red, 0; green, 0; blue, 0 }  ][line width=0.75]    (10.93,-3.29) .. controls (6.95,-1.4) and (3.31,-0.3) .. (0,0) .. controls (3.31,0.3) and (6.95,1.4) .. (10.93,3.29)   ;
%Straight Lines [id:da7527162662680911] 
\draw [color={rgb, 255:red, 208; green, 2; blue, 27 }  ,draw opacity=1 ] [dash pattern={on 4.5pt off 4.5pt}]  (382.81,65.15) -- (388.19,51.95) ;
%Straight Lines [id:da3668493008866167] 
\draw [color={rgb, 255:red, 208; green, 2; blue, 27 }  ,draw opacity=1 ] [dash pattern={on 4.5pt off 4.5pt}]  (387.7,160.87) -- (383.29,144.12) ;
%Straight Lines [id:da8860808012789847] 
\draw [color={rgb, 255:red, 208; green, 2; blue, 27 }  ,draw opacity=1 ] [dash pattern={on 4.5pt off 4.5pt}]  (604.03,159.61) -- (609.42,145.39) ;
%Straight Lines [id:da011021940936073893] 
\draw [color={rgb, 255:red, 208; green, 2; blue, 27 }  ,draw opacity=1 ] [dash pattern={on 4.5pt off 4.5pt}]  (608.93,46.87) -- (602.57,33.66) ;
%Straight Lines [id:da22356758697364532] 
\draw    (339,30.11) .. controls (341.29,29.56) and (342.72,30.43) .. (343.27,32.72) .. controls (343.82,35.01) and (345.24,35.88) .. (347.53,35.33) .. controls (349.82,34.78) and (351.25,35.65) .. (351.8,37.94) .. controls (352.35,40.23) and (353.77,41.09) .. (356.06,40.54) .. controls (358.35,39.99) and (359.78,40.86) .. (360.33,43.15) .. controls (360.88,45.44) and (362.3,46.31) .. (364.59,45.76) .. controls (366.88,45.21) and (368.31,46.08) .. (368.86,48.37) .. controls (369.41,50.66) and (370.83,51.53) .. (373.12,50.98) .. controls (375.41,50.43) and (376.84,51.3) .. (377.39,53.59) .. controls (377.94,55.88) and (379.36,56.75) .. (381.65,56.2) -- (385.5,58.55) -- (385.5,58.55) ;
%Straight Lines [id:da6645894729508091] 
\draw    (340.22,169) .. controls (341.22,166.87) and (342.79,166.3) .. (344.92,167.29) .. controls (347.05,168.28) and (348.62,167.71) .. (349.62,165.58) .. controls (350.61,163.44) and (352.18,162.87) .. (354.32,163.86) .. controls (356.45,164.85) and (358.02,164.28) .. (359.01,162.15) .. controls (360.01,160.02) and (361.58,159.45) .. (363.71,160.44) .. controls (365.84,161.43) and (367.41,160.86) .. (368.41,158.73) .. controls (369.4,156.59) and (370.97,156.02) .. (373.11,157.01) .. controls (375.24,158) and (376.81,157.43) .. (377.8,155.3) .. controls (378.8,153.17) and (380.37,152.6) .. (382.5,153.59) -- (385.5,152.5) -- (385.5,152.5) ;
%Straight Lines [id:da4501677603244114] 
\draw    (606.73,152.5) .. controls (608.78,151.35) and (610.39,151.8) .. (611.54,153.85) .. controls (612.69,155.9) and (614.3,156.35) .. (616.35,155.2) .. controls (618.4,154.05) and (620.01,154.5) .. (621.17,156.55) .. controls (622.32,158.6) and (623.93,159.05) .. (625.98,157.9) .. controls (628.03,156.75) and (629.64,157.2) .. (630.8,159.25) .. controls (631.95,161.3) and (633.56,161.75) .. (635.61,160.6) .. controls (637.66,159.45) and (639.27,159.9) .. (640.43,161.95) .. controls (641.58,164) and (643.19,164.45) .. (645.24,163.3) .. controls (647.29,162.15) and (648.9,162.6) .. (650.05,164.65) -- (652,165.19) -- (652,165.19) ;
%Straight Lines [id:da6312255228607064] 
\draw    (605.75,40.27) .. controls (606.67,38.1) and (608.21,37.47) .. (610.38,38.39) .. controls (612.55,39.31) and (614.1,38.68) .. (615.01,36.51) .. controls (615.93,34.34) and (617.48,33.71) .. (619.65,34.63) .. controls (621.82,35.55) and (623.37,34.92) .. (624.28,32.75) .. controls (625.2,30.58) and (626.75,29.95) .. (628.92,30.87) .. controls (631.09,31.79) and (632.64,31.16) .. (633.55,28.99) .. controls (634.46,26.82) and (636.01,26.19) .. (638.18,27.11) .. controls (640.35,28.03) and (641.89,27.41) .. (642.82,25.24) .. controls (643.73,23.07) and (645.28,22.44) .. (647.45,23.36) -- (648.33,23) -- (648.33,23) ;

% Text Node
\draw (115.37,78.6) node [anchor=north west][inner sep=0.75pt]    {$L\sim E/\cT$};
% Text Node
\draw (453.41,82.8) node [anchor=north west][inner sep=0.75pt]    {$L\sim E/\cT$};

\end{tikzpicture}
    \caption{\small{Left: The Gross-Mende saddle with external legs that extend to infinity. Right: The external legs are cut-off and a graviton vertex is added to regulate the  extrinsic curvature of the saddle and study the gravitational amplitude.}}
    \label{GMS}
\end{figure}
Notice that the above calculation can only be trusted if the saddle is mildly curved ($L\sqrt{\cT}\gg1$). Therefore, by decreasing $L$ to $\sim 1/\sqrt{\cT}$ the calculation starts breaking down and we can estimate the largest trustable contribution to the high-energy gravitational amplitude to satisfy\footnote{We write an inequality because there could be contributions from other channels to the gravitational amplitude.}
\begin{align}
    \log\left(|\mathcal{A}|\right)\gtrsim -\widetilde{\mathcal{O}}\left(\frac{E}{\sqrt{\cT}}\right)\,.
\end{align}
A Taylor expansion of the exponential amplitude now proceeds in analogy to the discussion below \eqref{sphere} such that we can conclude that any weakly coupled string with tension $\cT$is expected to give rise to a higher-derivative correction as in \eqref{eq:Scorrweakstring}. Comparing to the prediction of the species scale we obtain the bound \eqref{tensionbound} for $p=1$.  

Let us now turn to higher-dimensional weakly coupled branes. Similar to weakly coupled strings we consider a brane to be weakly coupled if the extrinsic curvature of the brane configuration is small compared to the tension. In this case the higher-derivative corrections to the worldvolume theory on the brane are suppressed and we can trust the worldvolume theory. Consider such a weakly coupled  $p$-brane with co-dimension $d-p-1\geq1$ and tension $\cT$. To show $\cT\gtrsim \Lambda^{p+1}$ we proceed via contradiction. Therefore suppose there exists $\epsilon \ll 1$ such that $\cT = \epsilon^{p+1} \Lambda^{p+1}\ll \Lambda^{p+1}$.  Then, one can compactify the theory on a $(p-1)$-dimensional torus of radius $R$ such that 
\begin{equation}
    R^{p-1} = \frac{1}{\epsilon^p \Lambda^{p-1}}\,. 
\end{equation}
Since $\epsilon\ll 1$ the radius of the compactification $R\gg \Lambda^{-1}$ such that the compactification is within the regime of validity of the target space effective field theory which has a cutoff $\Lambda$. We can now wrap the brane on this $(p-1)$-torus to obtain an effective string in the extended directions. Notice that the extrinsic curvature of this brane configuration is set by the inverse radius of the torus. On the other hand we have
\begin{equation}
    \cT^{\frac{1}{p+1}} = \epsilon \Lambda \gg \epsilon^{\frac{p}{p-1}}\Lambda = R^{-1}\,,\quad  \text{for} \quad p<\infty
\end{equation}
such that the extrinsic curvature of this brane configuration is small compared to the brane tension. Hence we have a weakly coupled brane that yields a weakly coupled string upon dimensional reduction on the $(p-1)$-torus. The tension $\widetilde{\cT}$ of this string is given by 
\begin{equation}
 \widetilde{\cT} = \cT R^{p-1} = \epsilon^{p+1} \epsilon^{-p} \Lambda^2 = \epsilon \Lambda^2\,.
\end{equation}
Since $\epsilon \ll 1 $ we hence get a weakly coupled string with tension $\widetilde{\cT} \ll \Lambda^2$ in contradiction to our previous result. Therefore, our assumption $\cT\ll \Lambda^{p+1}$ was incorrect and we find that any weakly coupled $p$-brane satisfies \eqref{tensionbound} which proves Proposition~\ref{prop:tension}. 

Note that our argument for the Proposition~\ref{prop:tension} is an example of UV/IR connection in quantum gravity. The branes with a mass scale much smaller than the Planck mass can form well-controlled saddles in processes that involve energies much higher than their mass scale and therefore contribute to the high energy gravitational amplitudes. Moreover, as explained below the equation \eqref{GME}, a saddle of a given topology becomes less curved and more trustable at higher energies. In this way, these saddles are similar to black holes which are under better perturbative control in the semiclassical gravity when they are heavier. This is due to the fact that the heavier black holes have smaller horizon curvatures.

Let us close by considering the case of D$p$-branes in critical string theory. The tension  $\cT_p$ of these branes and the coupling $g_{p+1}$ to the RR-form $C_{p+1}$ are given in terms of the string scale $M_s$ and the string coupling $g_s$ as 
\begin{equation}
    \cT_{p} \sim \frac{M_s^{p+1}}{g_s}\,,\qquad g_{p+1} \sim g_s M_s^{p+2-d/2}\,. 
\end{equation}
Using these expression we find that for a D$p$-brane with $L^{p+1}\sim \cT_p$ the constraint \eqref{tensioncouplingweak} implies 
\begin{equation}
    g_s M_s^{p+4-d/2}<\left(\frac{M_s^{p+1}}{g_s}\right)^{1+\frac{2-d}{2(p+1)}}\qquad \Rightarrow \qquad g_s^{2+\frac{d-2}{2(p+1)}} < 1\,. 
\end{equation}
Hence in the weak-coupling limit for the critical string, $g_s\ll 1$, the D$p$-branes are indeed weakly coupled under the RR-forms. On the other hand, for small string coupling the quantum gravity cutoff is set by the string scale, i.e. $\Lambda\sim M_s$ such that for $g_s\ll 1$ the tension of the weakly coupled D$p$-branes satisfies $\cT \gg \Lambda^{p+1}$ such that our Proposition~\ref{prop:tension} is indeed satisfied.

\section{Spectrum of weakly coupled gravitational theories}\label{sec:spectrum}
In the previous section we obtained two central results of this work. First we showed that there exists an effective field theory with cutoff given by the species scale $\Lambda$. Second, we argued that the tension of any weakly coupled $p$-brane is constrained to be bigger than $\Lambda^{p+1}$. In this section we utilize these crucial results to constrain the spectrum of states in a weakly coupled theory of gravity. In section~\ref{sec:constraintlighttower} we first constrain the kind of tower of weakly coupled states with mass below $\Lambda$. We then show in section~\ref{sec:Hagedorn} that for energies $\Lambda\ll E\ll M_{\rm min}$ the density of one-particle state needs to grow exponentially in energy reproducing a Hagedorn-like behavior. In section~\ref{sec:higherdimBH} we present an argument that the existence of extra dimensions and their size is directly related to the growth of $\rho(E)$ above $M_{\rm min}$. Finally, in section~\ref{sec:ESC} we show how these results, that were extracted from the properties of gravitational amplitudes, provide bottom-up evidence for a central statement of the Emergent String Conjecture~\cite{Lee:2019oct}.
\subsection{Constraining towers lighter than \texorpdfstring{$\Lambda$}{Lambda}}\label{sec:constraintlighttower}
In the previous section we argued that there needs to exist an EFT that is valid up to energies $\Lambda$. Importantly, this is the EFT that calculates the entropy of all black holes in the theory. On the other hand, we know that any field couples universally to gravity and therefore corrects the gravitational action. For the sake of generality, we also allow the effective theory to include weakly coupled defects whose worldvolume theory is described by a weakly coupled field theory.\footnote{Defects, for which the worldvolume theory is strongly coupled, and that do not admit a Lagrangian description at low energies are not relevant for our discussion since they cannot provide a tower of weakly coupled excitations.} Closed configurations of these defects can lead to additional particle states in the theory. In the following, we collectively refer to the weakly coupled particles and the weakly coupled defect states of the EFT that is valid up to energy $\Lambda$ as weakly coupled states. Note that these are the only states that can perturbatively couple to gravity. On the other hand, strongly 
coupled states, even if stable and light, might not be described by the EFT. As an example of such states, consider BPS states in Calabi--Yau threefold compactifications of M-theory. For these theories there exist finite distance limits along which an effective divisor shrinks to zero size leading to a theory containing a tensionless string. In these limits there exist an infinite tower of BPS states with vanishing central charge. However, since these states are strongly-coupled and associated to a strongly coupled, tensionless strings they do not allow for a description in terms of a weakly coupled EFT.

In this section we wish to constrain the tower of such weakly coupled states with mass $m$ in the regime of validity of the EFT with cutoff $\Lambda$, i.e. states with $m<\Lambda$. More precisely we show the following 
{\proposition \label{prop:lighttower} Consider a $d$-dimensional theory of gravity with quantum gravity cutoff $\Lambda$. Then in the weak-coupling limit $\Lambda\ll M_{\rm pl,d}$ the only tower of weakly coupled states with mass below $\Lambda$ can be a KK-tower.}\\

Since we are interested in states with mass parametrically below $\Lambda$, such states allow for a description in terms of an EFT, which may be higher-dimensional. This means that there exists a description of the theory involving only finitely many fields and defects that collectively account for all the states in the theory with energies below $\Lambda$. These states can arise as one-particle states in the field theory, as bound states of the field excitations, or from the weakly coupled defects which includes states obtained by wrapping this defect along the compact dimensions.
Note that Proposition~\ref{prop:lighttower} is also satisfied by confining large-$N$ gauge theories which have a tower of confined states. In the large $N$-limit, the energy scale $\Lambda$ at which the higher-derivative corrections become important matches the mass scale of the 't Hooft string which is also the mass scale of the tower of confined states.

Let us assume we identified the effective theory that describes the theory of gravity up to $\Lambda$ as a $D$-dimensional EFT. As we showed in section~\ref{sec:boundtension} any weakly coupled $p$-brane has tension at or above the species scale $\cT\gtrsim \Lambda^{p+1}$. As a consequence these states cannot give rise a tower of states with mass below $\Lambda$. Still, upon compactification on a $p$-dimensional manifold of radius $R$ such $p$-branes may give rise to a tower of particle states in the $(D-p)$-dimensional theory with mass $m\sim \cT R^p$. Since the cutoff of the EFT sets the minimal length scale $R_{\rm min}\sim \Lambda^{-1}$ for which the effective theory can be described in terms of the $D$-dimensional theory, the mass of this tower of states is bounded by $m\gtrsim \Lambda$. Therefore also after dimensional reduction a defect theory cannot give rise to a tower of states describable within a gravitational EFT with cutoff $\Lambda$. In addition we now argue that there can be only a finite number of weakly coupled particles (including bound states) in the D-dimensional EFT at weak coupling. This EFT is defined in the following way: for every finite value of $\Lambda/M_{\rm pl,d}$ we consider the special EFT that describes all black holes in the theory. As we vary $\Lambda/M_{\rm pl,d}$ this defines a family of EFTs. The EFT at weak coupling is defined as the limit of this family as $\Lambda/M_{\rm pl,d}\rightarrow 0$.\footnote{The existence of a well-defined theory as we take $\Lambda/M_{\rm pl,d}\rightarrow 0$ is an assumption about how this limit is taken. For example, in string theory, this achieved by moving on an infinite geodesic in the moduli space. However, one can easily consider a non-geodesic path towards the asymptotic boundary of the moduli space (corresponding to $\Lambda/M_{\rm pl,d}\rightarrow 0$) that oscillates between different weak coupling descriptions. We would not consider such a trajectory in the space of theories a weak-coupling \textit{limit}.} Since by definition this EFT is valid up to energies $\Lambda$, its partition function has to be finite when the theory is put on a thermal circle with circumference $\beta> \Lambda^{-1}$ which we keep fixed in units of $\Lambda$ as we take the weak-coupling limit. Finiteness of the partition function implies that the number of weakly coupled states with masses below $\Lambda$, which we denote by $N$, is also finite since 
\begin{align}
    \mathcal{Z}(\beta)\geq\int_0^\Lambda dm ~e^{-\beta m} \rho(m)\geq e^{-\beta\Lambda} \int_0^\Lambda dm ~\rho(m)=e^{-\beta\Lambda}N\,.
\end{align}
In the first inequality we bound the full partition function from below by only considering the contribution of the single-particle states with mass below $\Lambda$. Therefore, assuming that $\mathcal{Z}$ is finite, we conclude that the number of particles is bounded from above as 
\begin{align}
    N\leq \mathcal{Z}(\beta)e^{\beta\Lambda}\,,
\end{align}
for any $\beta>\Lambda^{-1}$ and is hence finite.

This leaves us with a finite number of particles in the $D$-dimensional EFT. Still these states can acquire different momenta in the $D-d$ compact dimensions (KK modes). As is well known, the compactification of the $D$-dimensional theory on a $(D-d)$-dimensional manifold of size $R$ leads to a KK spectrum with mass $m_{\rm KK}\sim 1/R$ which in the limit $R\rightarrow \infty$ (corresponding to the weak-coupling limit $\Lambda/M_\text{pl,d} \to 0$) indeed leads to an infinite tower of states with mass below $\Lambda$. Therefore the only possible tower of weakly coupled states with mass below $\Lambda$ is a KK-tower as claimed in Proposition~\ref{prop:lighttower}. 

\subsection{Universality of Hagedorn behavior}\label{sec:Hagedorn}
Our analysis so far has revealed that for a theory of quantum gravity there are at least two important energy scales, the quantum gravity cutoff $\Lambda$ and the mass of the minimal black hole $M_{\rm min}$. In general these two energy scales do not coincide. In this section we want to argue that the behavior of the density of one-particle states for energies between $\Lambda$ and $M_{\rm min}$ has universal properties in any theory of gravity with weak-coupling limit. Our main claim is summarized in 
{\proposition \label{prop:Hagedorn} Consider a $d$-dimensional theory of gravity in asymptotically flat spacetime in its weak-coupling limit $\Lambda/M_{\rm pl,d}\rightarrow 0$ with $\Lambda$ the quantum gravity cutoff as defined in section~\ref{ssec:speciesscale} and let $M_{\rm min}$ be the energy scale of the particle/black hole transition. Then for energies $\Lambda \ll E \ll M_{\rm min}$ the density of one-particle states scales as $\log \rho(E) \sim \widetilde{\mathcal{O}}\left(E/\Lambda\right)$.
}\\

To prove this Proposition we first recall from section~\ref{sec:ExistenceEFT} that in the effective theory of gravity the shortest meaningful length scale is set by $\Lambda^{-1}$. On the other hand we can consider the partition function for a given system in our theory at temperature $T$ which admits a Euclidean path integral definition with a thermal circle of size $\beta=1/T$. Since the shortest length scale is set by $\Lambda^{-1}$ the Euclidean effective field theory is only valid for $\beta\gtrsim\Lambda^{-1}$. For effective theories of quantum gravity we therefore expect the temperature to be bounded from above by $T\lesssim \Lambda$. For black holes, which are described by this effective theory, this relation is indeed satisfied. For example for Schwarzschild black holes the temperature is inversely proportional to the horizon radius---independent of the dimension of the spacetime. For the smallest black hole with mass $M_{\rm min}$ we therefore have $ T(M_{\rm min}) \sim 1/r_{\rm min} = \Lambda$.

Let us now turn our attention to particle systems. In section~\ref{sec:ExistenceEFT} we argued that there exists an EFT description that is valid up to energies $\Lambda$ and showed in section~\ref{sec:constraintlighttower} that all towers of states below $\Lambda$ can be understood as KK-towers. Let us assume that the EFT, which describes the theory up to $\Lambda$, lives in $D$ dimensions. Consider a box of size $L\gg \Lambda^{-1}$ and energy $E$ in $D$ dimensions. We are interested in the weak gravitational limits where $L\Lambda$ and $E/\Lambda$ are kept fixed while $\Lambda/M_{\rm pl,D}\rightarrow0$. Since the $D$-dimensional theory obtained in the gravitational weak-coupling limit has no tower of particles with mass below $\Lambda$, the mass $m_{\rm max}$ of the heaviest state in this EFT is much smaller than $\Lambda$, $m_{\rm max}\ll \Lambda$. By inserting energy into the box we increase the temperature of the system in the box above $m_{\rm max}$ such that the light particles in the EFT can be treated as effectively massless. Therefore, for temperatures $m_{\rm max}\ll T<\Lambda$, we have 
\begin{equation}
    E\sim T^D L^{D-1},
\end{equation}
which is the energy of radiation at temperature $T$ inside a $(D-1)$-dimensional spatial box. At energy $E_{c}=\Lambda^D L^{D-1}$ the temperature will be proportional to $ \Lambda$. Importantly, in the gravitational weak-coupling limit we keep $L\Lambda$ fixed such that $E_c\propto \Lambda$. Therefore, this energy will be much smaller than the mass of a black hole of the same size as the box (i.e. radius $L$) which is given by $E_{\max}=L^{D-3}M_{\rm pl, D}^{D-2}$. Since $E_c\ll E_{\max}$ the system has not collapsed into black hole and hence is indeed gravitationally weakly coupled. 

At $E_c$ the temperature is already of the order of the maximal temperature $T_{\max}\propto \Lambda$ such that increasing the energy beyond $E_c$ cannot lead to a further parametric increase of the temperature. And indeed, from Proposition~\ref{propparticleBH} we know that below $M_{\rm min}$, $\Omega(E)$ cannot grow super-exponentially, i.e. $\Omega(E)$ is given by \eqref{eq:Omega-general} with $\alpha_\Omega\leq 1$. For energies $E_c<E\ll E_{\max}$ the temperature in the microcanonical ensemble thus remains essentially constant:
\begin{align}
    T(E)^{-1}=\frac{{\rm d}\log\Omega}{{\rm d}E}\sim \widetilde{\mathcal{O}}\left(\Lambda^{-1}\right)\,,\quad \text{for}\quad E_c<E\ll E_{\max}\,.
\end{align}
Integrating the above equation leads to
\begin{align}
    \log\Omega(E)\propto \widetilde{\mathcal{O}}\left( \frac{E}{\Lambda}\right)\,, \quad \text{for}\quad E_c<E\ll E_{\max}\,.
\end{align}
Comparing with \eqref{eq:Omega-general} this implies $\alpha_\Omega=1$ in the energy window above. Suppose the average energy of \emph{single} particles near the upper end of the energy interval is given by $\epsilon_{\max}\gg \Lambda$. On the other hand, since at the lower end of the energy interval we are dealing with radiation in a box of temperature $\mathcal{O}(\Lambda)$, the average energy of a single particle at the lower end the interval $\epsilon_{\rm min}$ is proportional to $\Lambda$, $\epsilon_{\min} \propto \Lambda$. In the weak-coupling limit $\epsilon_{\max}\rightarrow\infty$ though its exact value in terms of $\Lambda$ and $M_{\rm pl,D}$ in this limit depends on the profile of $\rho(E)$. From \eqref{uin2} we recall that an exponential growth of $\Omega$ implies an exponential growth for $\rho(E)$. This, in turn, implies
\begin{align}
        \log \rho(E)\sim \widetilde{\mathcal{O}}\left(\frac{E}{\Lambda}\right)\,,\quad \text{for}\quad \Lambda\ll E\ll \epsilon_{\max}\,.
\end{align}
We thus conclude that the density of states $\rho(E)$ at energy $E=x\Lambda$ converges to an exponential function in $E/\Lambda$ in the weak-coupling limit for any $x\gg1$. The very fact that the density of particles converges means that  in the gravitational weak-coupling limit there is a tower of states with exponentially growing density for which the masses are proportional to $\Lambda$.\footnote{In string theory, these states are the string excitations and $\Lambda$ is the string scale.} 

To argue for a tower of states with masses proportional to $\Lambda$ and exponential density in the energy window $\Lambda\ll E\ll M_{\rm min}$, consider a state for which the mass converges to $x\Lambda$ in the weak-coupling limit for some $x\gg 1$. If we move away from the weak-coupling limit, the gravitational coupling for this state increases until the gravitational self-energy becomes dominant and it collapses into a black hole. However, as long as we are at a point along the weak-coupling limit where the mass of the state is much smaller than the mass, $M_{\rm min}$, of the smallest black hole, the gravitational corrections to the mass of the state are small. Therefore for energies $\Lambda\ll E\ll M_{\rm min}$ there indeed exist states with exponential density and mass proportional to $\Lambda$. Note that the tower we considered here might not include all particles. For example, there could be a tower of particles for which the mass in the weak-coupling limit is not bounded in units of $\Lambda$.\footnote{An example of such tower is $D0$ branes in type IIA string theory in the limit where the string coupling goes to zero.} Still, we showed that there exists a subset of particles with mass scale $\Lambda$ with exponentially growing density $\log \rho(E)\propto E/\Lambda$ up to logarithmic corrections in $E/\Lambda$. In fact the contribution to $\rho$ from other states has to be of the same order or subleading. If this was not the case, the number of one-particle states at energy $E=M_{\rm min}$ would be parametrically larger than $\exp(M_{\rm min}/\Lambda)$ which is the density of states of the smallest black hole with entropy $M_{\rm min}/\Lambda$. Altogether we can hence conclude that the tower of states with masses between $\Lambda$ and $M_{\rm min}$ leads to $\log \rho(E) \propto E/\Lambda$ up to logarithmic corrections as claimed in Proposition~\ref{prop:Hagedorn}.

Note that our results are consistent with the existence of the Horowitz--Polchinski solution~\cite{Horowitz:1997jc} as an additional phase before reaching $M_{\rm min}$. These solutions describe selfgravitating string backgrounds and their entropy to leading order is linear in energy~\cite{Chen:2021dsw}. Therefore, they satisfy Proposition~\ref{prop:Hagedorn}.

\subsection{High-energy signature of extra dimensions}\label{sec:higherdimBH}
In section~\ref{sec:particleBH} we argued that the states at energy $E$ are black hole microstates if and only if $\log \rho(E) \sim (E/\Lambda)^\alpha$ for $\alpha>1$ (see Proposition~\ref{propparticleBH}). In particular we saw that in a $d$-dimensional theory of gravity Schwarzschild black holes lead to $\alpha =\frac{d-2}{d-3} > 1$. In this section, we show that the weak energy condition at low energies implies that the values for $\alpha$ in the range $1<\alpha<\frac{d-2}{d-3}$ can only be attained in the presence of extra dimensions.

{\proposition \label{prop:higherdimBH} Consider a $d$-dimensional theory of gravity in asymptotically flat spacetime with quantum gravity cutoff $\Lambda$. In theories that the weak energy condition is satisfied by the low energy field theory, the density, $\rho$, of one-particle states as a function of energy, $E$, can grow like $\log \rho(E) \propto (E/\Lambda)^{\alpha}$ with $1<\alpha<\frac{d-2}{d-3}$ only in the presence of extra dimensions. The one-particle states leading to such a growth of $\rho$ are the microstates of black holes localized in the compact dimensions with horizon size smaller than the size of the extra dimensions.}\\

Notice that for large energies $\alpha=\frac{d-2}{d-3}$ since for any compact space there exists a mass threshold beyond which the horizon of any black hole is larger than the size of this compact space. Therefore for large energies all black holes will be $d$-dimensional. Still at lower energies the effective scaling of $\rho(E)$ parametrized by $\alpha$ can change. In order to show that a change from $\alpha=\frac{d-2}{d-3}$ to $1<\alpha<\frac{d-2}{d-3}$ implies the presence of extra dimensions, we first need to argue such a change cannot be caused by adding some arbitrary but finite number of states to the theory.

To that end we use the weak energy condition (which is expected to hold in asymptotically flat spacetimes) to show that for a fixed energy one cannot construct a state with entropy higher than the Schwarzschild black hole by adding a finite number of arbitrary fields and interactions to the theory. To see that consider a $d$-dimensional spherically symmetric black hole with fixed mass $M$ described by a corrected version of the Schwarzschild metric for which the corrections arise due to the various gravitational couplings in the action. Using the weak energy condition, we show that the corrections decrease the entropy of the black hole. Suppose the metric is given by
\begin{align}\label{correctedmetric}
ds^2=-e^{2\nu(r)}dt^2+e^{2\lambda(r)}dr^2+r^2d\Omega_{d-2}^2\,,
\end{align}
where $\nu(r)$ and $\lambda(r)$ are functions encoding the corrections to the Schwarzschild metric. The uncorrected Schwarzschild solution corresponds to 
\begin{equation}\label{lambdas}
    e^{-2 \lambda_S(r)}=e^{2\nu_S(r)} = 1-\frac{\kappa M}{4\pi r^{d-3}}\,,
\end{equation}
where $\kappa = M_{\rm pl,d}^{2-d}$. The entropy of neutral large black holes at leading order is given by $S = \frac{A}{4G} +\mathcal{O}(\log A)$~\cite{Sen:2012dw}. Therefore, to show that the corrections to the Schwarzschild metric in \eqref{correctedmetric} decrease the entropy, we only need to show that the radius of the black hole described by \eqref{correctedmetric} is smaller than the radius of a Schwarzschild black hole with the same mass.

We can define coordinates such that $\lim_{r\rightarrow\infty}\nu(r)=0$ which, via the Einstein equations, also implies $\lim_{r\rightarrow\infty}\lambda(r)=0$. Suppose the horizon radius is given by $r_{H}$ and the mass of the ADM black hole is given by $M$. Then we have 
\begin{align}
    \lim_{r\rightarrow r_{H}}e^{-\lambda(r)}=\lim_{r\rightarrow r_{ H}}e^{\nu(r)}=0\,,
\end{align}
and
\begin{align}
    \lim_{r\rightarrow \infty}(1-e^{2\nu(r)})r^{d-3}=\lim_{r\rightarrow \infty}(1-e^{-2\lambda(r)})r^{d-3}=\frac{\kappa M}{4\pi}\,.
    \label{eq:ADM-mass}
\end{align}
Note that the relationship between $M$ and $r_{H}$ is no longer $r_{H}^{d-3}=\kappa M/(4\pi)$ due to the corrections. The Einstein equations give 
\begin{equation}
\kappa T^0_0=-(d-2) \frac{e^{-2\lambda}}{r^2}\left[\frac{(d-3)}{2}(-1+e^{2\lambda})+ \,r\lambda'\right]\,.
\end{equation}
where $'$ denotes derivatives with respect to $r$. On the other hand, the weak energy condition implies 
\begin{equation}
T^0_0\leq 0\,.
\end{equation}
Hence, we find 
\begin{align}
    \frac{(d-3)}{2}\left(-1+e^{2\lambda}\right)+ \,r\lambda'&\geq 0\:.
\end{align}
The above inequality is saturated for the Schwarzschild solution with $\lambda=\lambda_S$ defined in \eqref{lambdas}. We therefore find
\begin{align}\label{SMR}
    \lambda'-\lambda'_S\geq \frac{(d-3)}{2}\frac{e^{2\lambda_S}-e^{2\lambda}}{r}\,.
\end{align}
We now show that for a black hole with ADM mass given in \eqref{eq:ADM-mass} the above inequality implies $e^{2\lambda}\leq e^{2\lambda_S}$ where $\lambda_S$ corresponds to the metric of a Schwarzschild metric of the \emph{same} mass. 
Since we fix the ADM mass $M$, for large $r$ we have
\begin{align}
    e^{2\lambda} &= 1+ \frac{\kappa M}{4\pi r^{d-3}} + \mathcal{O}(r^{6-2d}) \:,\qquad
    e^{2\lambda_S} = 1+ \frac{\kappa M}{4 \pi r^{d-3}} + \mathcal{O}(r^{6-2d})\:,
    \label{eq:LambdaExpand}
\end{align}
such that the difference between $e^{2\lambda}$ and $e^{2\lambda_S}$ is $\mathcal{O}(r^{6-2d})$. In terms of
\begin{equation}
    \delta\lambda \equiv \lambda - \lambda_S\,,
    \label{eq:deltaLambda}
\end{equation}
we can rewrite the inequality \eqref{SMR} in the $r\to \infty$ limit as
\begin{equation}
    (\delta \lambda)' \geq \frac{(d-3)}{2r} \left( 1+ \frac{\kappa M}{4 \pi r^{d-3}} + \mathcal{O}(r^{6-2d}) \right) \left(-2\delta \lambda + \mathcal{O}(\delta \lambda)^2\right)\,.
\end{equation}
Keeping only the leading orders in $1/r$ and $\delta\lambda$ we find 
\begin{equation}\label{leadingSMR}
   (\delta \lambda)'\geq -\frac{(d-3)}{r} \delta \lambda\,.
\end{equation}
This form of the inequality suggests the ansatz
\begin{equation}
    \delta\lambda = \frac{f}{r^{d-3}}\,,
\end{equation}
for some function $f=f(r)$. Via \eqref{eq:LambdaExpand} the requirement that the ADM mass of the corrected black hole is equal to the mass of a Schwarzschild black hole translates into
\begin{equation}
    \lim_{r\to \infty} f = 0\:.
\end{equation}
From \eqref{leadingSMR} we further find that at large radius
\begin{equation}
    f' \geq 0\:.
\end{equation}
These conditions on $f$ imply $\delta$ approaches zero from below as $r\to \infty$, which means $e^{2\lambda} \leq e^{2\lambda_S}$ for sufficiently large $r$. The condition \eqref{SMR} then translates to $0>\lambda'\geq\lambda'_S$ and $e^{2\lambda}\leq e^{2\lambda_S}$ everywhere. Since $e^{2\lambda}\leq e^{2\lambda_S}$, $\lambda$ is finite whenever $\lambda_S$ is finite. Therefore, $\lambda(r)$ is finite for all radii $r\geq r_S(M)=(\kappa M/(4\pi))^{1/(d-3)}$. Hence, we find that the horizon radius $r_{H}$ at which $\lambda$ diverges satisfies $r_{H}\leq r_S$, which is the desired result.

In order to obtain a black object whose entropy exceeds that of a Schwarzschild black hole of the same mass, we need to add an \emph{infinite} number of states to the theory. Such states can be provided by an infinite tower of states. However, as we argued in section~\ref{sec:ExistenceEFT}, black holes with mass $E>M_{\rm min}$ have to be describable in terms of \textit{some} EFT. From Proposition~\ref{prop:lighttower} we recall that the only infinite tower of weakly coupled states with mass below $\Lambda$ can be a KK-tower arising from dimensional reduction of a higher-dimensional theory. The objects with entropy higher than a $d$-dimensional Schwarzschild black hole (at fixed energy) are therefore black holes in higher dimensions.

Concretely, consider a $D$-dimensional theory compactified on a $(D-d)$-dimensional manifold.  For simplicity, let us consider a homogeneous internal geometry characterized by a length scale $R$ and let $M_{\rm pl,D}$ and $M_{\rm pl,d}$ be, respectively, the $D$- and $d$-dimensional Planck scales which are related via 
\begin{align}\label{eq:Mpl}
    M_{{\rm pl,D}}^{D-2}=M_{{\rm pl,d}}^{d-2}R^{d-D}.
\end{align}
In this construction, $D$-dimensional black holes correspond to localized objects inside the compact geometry which, however, only exist for $r_{H}<R$. If we increase the energy and therefore the horizon radius of this black hole to $r_{ H} > R$, the black hole eventually reaches the size of the compact geometry and the backreaction causes the horizon to form a black brane wrapping the internal geometry. In the lower dimensional theory, this black brane corresponds to a black hole. The condition $r_{H} \geq R$ translates to  
\begin{align}\label{eq:MC}
E \geq m_c \equiv R^{d-3}M_{{\rm pl,d}}^{d-2}\,.
\end{align} 
Note that for $E \leq m_c$, the black hole and the black brane solutions are both sensible classical solutions, but the black brane solution has higher entropy and is more stable thermodynamically. Consider for example the entropy of the localized $D$-dimensional and the $d$-dimensional Schwarzschild black holes in the microcanonical ensemble
\begin{equation}
\begin{aligned}\label{eq:higherdimdominance}
    S_{\rm BH,d}(E)\sim\left(\frac{E}{M_{{\rm pl,d}}}\right)^\frac{d-2}{d-3} 
    &= \left(\frac{E}{m_c}\right)^{\frac{d-2}{d-3}} (R M_{{\rm pl,d}})^{d-2}\:,
    \\
    S_{\rm BH,D}(E)\sim\left(\frac{E}{M_{{\rm pl,D}}}\right)^\frac{D-2}{D-3} 
    &= \left(\frac{E}{m_c}\right)^{\frac{D-2}{D-3}} (R M_{{\rm pl,d}})^{d-2}\:,
\end{aligned}
\end{equation}
where in the last equalities we used~\eqref{eq:Mpl} and~\eqref{eq:MC}. At $E\gg m_{c}$ the most entropic states are the $d$-dimensional Schwarzschild black holes. These, however, become unstable at energies $E\sim m_c$ corresponding to the Gregory--Laflamme transition~\cite{Gregory:1993vy}. At energies $E\ll m_c$ the entropically favored black hole is the $D$-dimensional one, as is clear from \eqref{eq:higherdimdominance} since $\frac{D-2}{D-3} < \frac{d-2}{d-3}$. As a consequence, the smallest-sized black hole trustable in \emph{any} effective theory is $D$-dimensional black hole, with $D$ the highest possible dimension. 
We see that, indeed, $1<\alpha<\frac{d-2}{d-3}$ can be achieved in the presence of extra dimensions at energies $E$ corresponding to black holes with a horizon radius smaller than the extra dimensions. According to our previous discussion this is also the only way to obtain such a value for $\alpha$ which is the statement of Proposition~\ref{prop:higherdimBH}. 

Let us close with some comments about the above result:
\begin{enumerate}
\item If both the lower- and the higher-dimensional black holes contributed to the amplitude as $\exp(-\mathcal{O}(S))$, where $S$ is their respective entropy, the contribution from the  black hole with \textit{smaller} entropy would dominate. This would undermine the claim that the gravitational amplitude captures the more stable saddle. However, this does not happen. Suppose we are in an energy window in which the higher- and the lower-dimensional black hole solutions are both well-defined, but the lower-dimensional black hole is unstable due to the Gregory--Laflamme instability~\cite{Gregory:1993vy}. The inverse of the growth rate of the tachyonic mode which sets the time scale for this transition is $\sim r_H$. This time scale is much smaller than the lifetime $\sim r_H^{d-1}M_{\rm pl, d}^{d-2}$ set by the Hawking radiation. Therefore, the black hole transforms into the more stable higher-dimensional configuration before Hawking radiating into two particles. This was expected since the Gregory--Laflamme instability would only be meaningful if the decay time between the two black holes is short enough to be seen before the black hole evaporates. 

\item The scale $m_c$ that occurs in the context of extra compact dimensions is an example of a more general scale, the black hole scale $\Lambda_{\rm BH}$, that was recently conjectured to be universally present in theories of quantum gravity where the black hole undergoes a phase transition to a entropically favored solution~\cite{Bedroya:2024uva}. As argued in there, in the case of a string tower, the scale $\Lambda_{\rm BH}$ is identified with the scale of the Horowitz-Polchinski transition. 

\item Finally, we note that an important consequence of our results is that the higher-dimensional black hole cannot be dimensionally reduced. In other words, no EFT with finitely many KK particles can approximate the higher-dimensional black hole and its entropy. Instead, the higher-dimensional black hole must be understood as a condensate of infinitely many KK states. Our argument provides a sharp explanation for the top--down observation made in~\cite{Bedroya:2024uva} that a thermodynamic transition of the black hole to a more stable phase is correlated with the existence of an infinite tower of states. The known examples include the Gregory--Laflamme transition which is accompanied by the KK tower and the the Horowitz--Polchinski solution~\cite{Horowitz:1997jc} in $d<7$ which is accompanied by a tower of string excitations.
\end{enumerate}
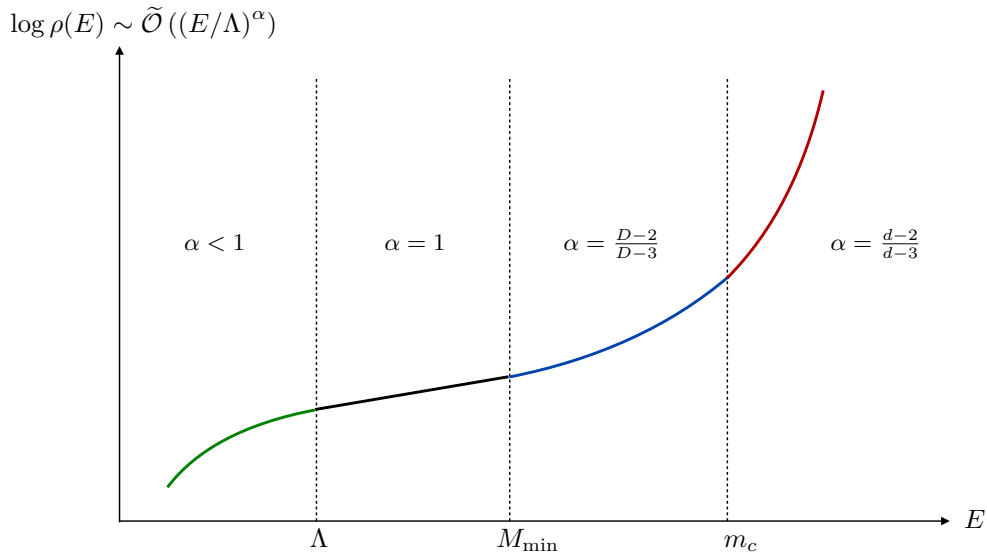
\begin{figure}[t]
\begin{center}
\includegraphics[width=0.75\textwidth]{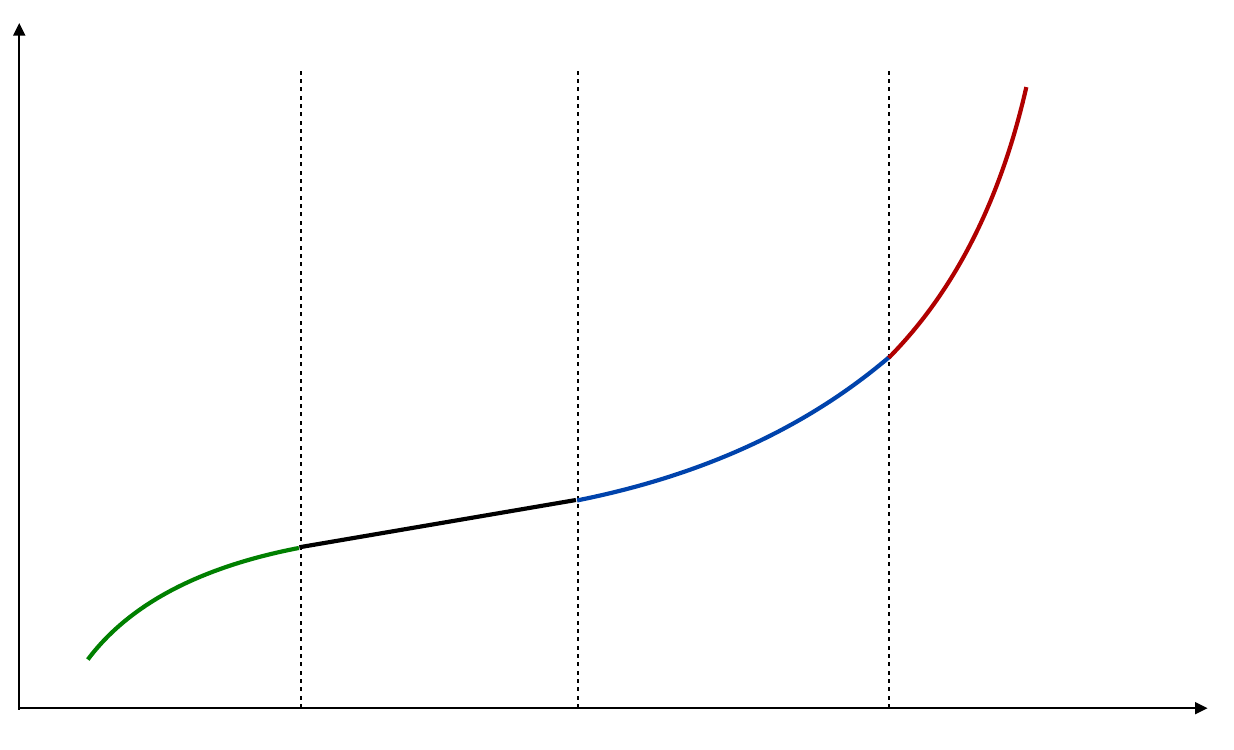}
\end{center}
\begin{picture}(0,0)
\put(375,35){\small $E$}
\put(130,28){\small $\Lambda$}
\put(200,28){\small $M_{\rm min}$}
\put(285,28){\small $m_{c}$}
\put(18,222){\small $\log \rho(E)\sim \widetilde{\mathcal{O}}\left(\left(E/\Lambda\right)^\alpha\right)$}
\put(83,140){\footnotesize $\alpha<1$}
\put(158,140){\footnotesize $\alpha=1$}
\put(225,140){\footnotesize $\alpha=\frac{D-2}{D-3}$}
\put(325,140){\footnotesize $\alpha=\frac{d-2}{d-3}$}
\end{picture}
\caption{\label{fig:doe}\small{The density of one-particle states $\rho$ as a function of energy $\log \rho(E) \sim \widetilde{\mathcal{O}}(E^\alpha)$ in a $D$-dimensional theory with $(D-d)$ compact dimensions of radius $R$ in the weak-coupling limit $\Lambda\ll M_{\rm pl,d}$. According to the Propositions~\ref{propparticleBH}, \ref{prop:Hagedorn} and \ref{prop:higherdimBH} the parametric scaling of $\log \rho(E)$ is universal in the energy windows set by the quantum gravity cutoff, $\Lambda$, the mass of the minimal black hole $M_{\rm min}$ and the mass scale $m_c$ in \eqref{eq:MC}.}}
\end{figure}
\subsection{Relation to Emergent String Conjecture}\label{sec:ESC}
The results of the previous sections provide strong constraints on the spectrum of massive one-particle states in theories of quantum gravity. In particular the Propositions~\ref{propparticleBH}, \ref{prop:Hagedorn}, and~\ref{prop:higherdimBH} constrain the scaling of the density of one-particle states $\rho$ as a function of energy in the gravitational weak-coupling limits $\Lambda \ll M_{\rm pl}$. A sketch of this behavior is provided in Figure~\ref{fig:doe}. 

Together with Proposition~\ref{prop:lighttower} these results nicely relate to the Emergent String Conjecture (ESC)~\cite{Lee:2019oct} which, as reviewed in the introduction, states that the lightest tower of states in any asymptotic limit of quantum gravity is either a KK-tower or consists of the excitations of a fundamental string. In light of the ESC, it is important to notice that our results as summarized in Propositions~\ref{propparticleBH}-\ref{prop:higherdimBH} have a built-in UV-IR connection that allows us to extract the properties of towers of light states from the behavior of $\rho$ at much higher energies. It follows from Propositions~\ref{propparticleBH} and \ref{prop:Hagedorn} that we can use $\rho(E)$ to read off $\Lambda$ and $M_{\rm min}$. Working from now on in $d$-dimensional Planck units, we can differentiate the cases\footnote{By Proposition \ref{prop:Hagedorn} the third possibility $M_{\rm min} \gg \Lambda^{3-d}$ (in Planck units) can never be realized since the exponential degeneracy in the interval $[\Lambda, M_{\rm min}]$ causes states to form black holes at energies above $\Lambda^{3-d}$.}
\begin{equation}
  a):\; M_{\rm min} \ll \Lambda^{3-d}\,,\quad b):\;M_{\rm min} \sim \Lambda^{3-d}\:.
\end{equation}
For a Schwarzschild black hole in $d$ dimensions its mass and temperature are related via 
\begin{equation}\label{eq:MofT}
 M_{\rm BH,d} \sim T^{3-d}\,.
\end{equation} 
Recall that the minimal mass black hole with mass $M_{\rm min}$ has temperature $T=\Lambda$. In case $a)$ this minimal mass black hole therefore cannot be $d$-dimensional, and needs to be a different state which is entropically favored over the $d$-dimensional Schwarzschild black hole. By Proposition~\ref{prop:higherdimBH} this has to be due to the existence of $(D-d)\geq 1$ extra dimensions with radius $R$ such that the smallest mass black hole is a $D$-dimensional black hole with mass (in $d$-dimensional Planck units)
\begin{equation}
M_{\rm min} = \frac{R^{d-3}}{(\Lambda R)^{D-3}}\,. 
\end{equation} 
Since this black hole only exists for energies $E<m_c$, with $m_c$ defined in \eqref{eq:MC}, we need to have $M_{\rm min}<m_c$ leading to the constraint
\begin{equation}
R^{-1}<\Lambda\,,\qquad\Leftrightarrow \qquad m_{\rm KK} < \Lambda\,, 
\end{equation}
implying that there is a KK-tower with mass below $\Lambda$. By Proposition~\ref{prop:lighttower} this KK-tower has to be the lightest tower of states since there cannot exist any other tower with mass below $\Lambda$. Case $a)$ is the situation depicted in Figure~\ref{fig:doe}. On the other hand in case $b)$ the minimal mass black hole \textit{is} a $d$-dimensional black hole. Even in the presence of extra dimensions, $m_c\leq M_{\rm min}$ and higher-dimensional black holes do not exist. Accordingly there is no KK tower with mass below $\Lambda$.

Due to the presence of a tower of light states, infinite distance limits in field space precisely correspond to weak-coupling limits for which $\Lambda\ll M_{\rm pl}$. Hence our results concerning the properties of the density of one-particle states apply in particular to these limits.  The statements of Proposition~\ref{prop:lighttower} and \ref{prop:Hagedorn} can then be reinterpreted as bottom-up evidence for the ESC since they can be reformulated into the following

{\corollary \label{mESC}
  In any weak-coupling limit ($\Lambda\ll M_{\rm pl}$) of an Einstein theory of gravity in asymptotically flat space, the lightest tower of weakly coupled states is either $a)$ 
   a KK-tower, or $b)$ a tower of states with exponential density of states $\rho(E) \sim e^{E/\Lambda}$. 
}\\

Notice that fundamental strings predicted by the ESC precisely have degeneracies that grow exponentially in the energy level $\log \rho(E) \sim E/M_s$, and for fundamental strings the species scale is given by the string scale $\Lambda=M_s$. The statement in Corollary~\ref{mESC} is agnostic about the existence of a duality frame in which this tower arises from a perturbative fundamental string. Still our result implies that (in asymptotically flat space) there are essentially just two options for the structure of the light states in quantum gravity. Even though in case the tower is not a KK-tower we cannot show that the states arise from fundamental strings, our results imply that in this situation the Hagedorn-growth for these states is universal.

Let us close by noticing that by \eqref{OmegaKKstates} an exponentially growing $\rho(E)$ can be viewed as the limit $p\rightarrow \infty$ for the density of KK-states on a $p$-dimensional compact manifold. Corollary~\ref{mESC} therefore implies that the lightest tower of states either behaves like a KK-tower for finite $p$ or like a KK-tower in the limit $p\rightarrow \infty$. This is interesting since the $\mathcal{O}(1)$ coefficients appearing in the Distance Conjecture are bounded by the values obtained for emergent string limits~\cite{Etheredge:2022opl,vandeHeisteeg:2023ubh,vandeHeisteeg:2023dlw} which always correspond to the $p\rightarrow \infty$ limit for the values of decompactification limits. As our results imply that the most extreme density of a light tower of states can be viewed as the $p\rightarrow \infty$ limit for a KK-tower this provides further evidence for the bounds on $\mathcal{O}(1)$ parameters derived from the Emergent String Conjecture to hold in general.

\section{Discussion}\label{sec:discussion}
In this work we investigated universal features of the density, $\rho$, of one-particle states in theories of quantum gravity. More precisely we studied the universal behavior of $\rho$ as a function of energy in weak-coupling limits in which the quantum gravity cutoff can be made parametrically smaller than the Planck scale, $\Lambda\ll M_{\rm pl}$. Central to our analysis was our estimate of the density of one-particle states $\rho$ discussed in section~\ref{sec:density}. The behavior of scattering amplitudes in perturbative string theory and the black hole regime served as the motivation to estimate $\rho$ via scattering amplitudes for energies above the cutoff $\Lambda$. Let us stress that we did not provide a first-principle derivation of the identification of these two quantities. It would be interesting to further scrutinize this identification in future work. 

Scattering amplitudes played a crucial role in clarifying the appearance of the EFT cutoff in the higher-derivative terms correcting the Einstein--Hilbert action. Even though it has been argued previously~\cite{vandeHeisteeg:2023dlw,vandeHeisteeg:2023ubh} that the species scale $\Lambda$ should be the scale suppressing these higher-derivative terms in the effective action, the equivalence of this scale to the EFT cutoff set by the radius of the smallest black holes describable in the EFT had not been demonstrated explicitly. In this work, we provided an argument for the identification between this equivalence based on the properties of gravitational scattering amplitudes. In particular our argument crucially made use of the features of scattering amplitudes in the black hole regime thereby highlighting the gravitational nature of the effective theory. We believe that our approach of exploiting gravitational scattering amplitudes to investigate important quantities for recent developments within the Swampland program (in this case the species scale) is a promising avenue to further pursue.

Gravitational scattering amplitudes in combination with effective string theory~\cite{Polchinski:1991ax} further led to the bound on the tension of weakly coupled $p$-branes in terms of the species scale. Whereas such a bound has previously been observed in explicit string theory compactifications (see, e.g.,~\cite{Cota:2022yjw}) our result gives the first bottom-up argument for such a bound. In the context of our analysis this result proved useful in arguing that the only tower of weakly coupled states below $\Lambda$ can be a KK-tower indicating strong constraints on such possible towers in an EFT. On the other hand, at energies $\Lambda\ll E\ll M_{\rm min}$ we showed that the density of one-particle states scales exponentially in the energy leading to an approximately constant temperature in that energy window, up to logarithmic corrections. From the perspective of the Swampland program, our result provides evidence for the Emergent String Conjecture~\cite{Lee:2019oct} by showing that in asymptotically flat spacetime the lightest tower of states has to either be a KK tower or a tower with Hagedorn growth for the one-particle states, a behavior familiar from fundamental strings. Let us stress that with our argument we are not able to infer the exact origin of the tower of states with exponential degeneracy, i.e., whether the states are always excitations of fundamental strings as predicted by the ESC. It would be interesting to see whether one can argue for this feature of the tower utilizing only bottom-up arguments. Even in explicit string theory constructions the identification of the fundamental string requires detailed knowledge of the intricate dualities of string theory. Hence, addressing this aspect of the ESC requires a bottom-up derivation of string dualities, see e.g.~\cite{Bedroya:2023xue} for a recent discussion of deriving string dualities using Swampland principles. 

Still, since the ESC is at the bedrock of many other arguments for non-trivial, universal features of quantum gravity, we believe that our result provides an important step towards consolidating the validity of the ESC. At its heart the ESC quantifies the observation in string theory, that the leading tower of light states in quantum gravity always collectively describes a richer semi-classical theory. The fact that in this work we are able to provide bottom-up evidence for one of the prediction of the ESC can be understood as a consequence of the emergence of such a semi-classical description.  Another semi-classical signature of quantum gravity is the validity of the Euclidean Path Integral (EPI) in uncovering the UV properties of gravity. In fact we used the validity of the EPI description in our analysis in section~\ref{sec:spectrum}, when identifying the maximal temperature describable within the EFT with the quantum gravity cutoff, and when discussing the effect of additional fields on corrections to the entropy of Schwarzschild black holes. It would be interesting to explore the relation between the semi-classical description of gravity predicted by the ESC and the EPI in more detail. Another important aspect of the ESC that we have not addressed in this work is the prediction that KK towers or towers of excitations of fundamental string necessarily have to arise at infinite distance in the scalar field space. In other words, we have not addressed the part of the ESC corresponding to the Distance Conjecture. It would be very interesting to see whether methods akin to those exploited in this work can also provide bottom-up evidence for the Distance Conjecture itself, see \cite{Calderon-Infante:2023ler} for a related discussion.

In any event, we believe that our results demonstrate the significance of combining different perspectives on quantum gravity. For example our work suggests a natural connection to the amplitude bootstrap program, given our non-trivial results for the density of massive states in quantum gravity. Even though our considerations here have been for black holes in asymptotically flat spacetimes, we can envision extending some of the arguments to small black holes in asymptotically Anti-de Sitter space, where such black holes are not eternal due to Hawking radiation, and are qualitatively similar to the flat case. In this way it might also be possible to provide bottom-up evidence for Swampland constraints without assuming asymptotically flat spacetimes. 

\subsubsection*{Acknowledgments}
We have greatly benefited from discussions with Nima Arkani-Hamed, Alberto Castellano, Eduardo Garcia-Valdecasas, Damian van de Heisteeg, Dieter Lüst, John Stout, and Cumrun Vafa. We are especially grateful to Matt Reece for comments on the draft. The work of AB is supported in part by the Simons Foundation grant number 654561 and by the Princeton Gravity Initiative at Princeton University. The work of MW is supported by a grant from the Simons Foundation (602883,CV), the DellaPietra Foundation, and by the NSF grant PHY-2013858.  The work of RKM is supported by NSF grants PHY-1620806, PHY-1748958 and PHY-1915071, the Chau Foundation HS Chau postdoc award, the Kavli Foundation grant ``Kavli Dream Team'', and the Moore Foundation Award 8342.
\appendix
\section{Effective strings}\label{A1}

In this section we review the theory of effective strings which is the framework that studies non-fundamental strings~\cite{Polchinski:1991ax}. Let us start by explaining precisely what is different between a generic effective string, and the fundamental string that is quantized in string theory. In the following we focus for simplicity on bosonic strings.

The common meaning of a string is a $(1+1)$-dimensional defect that couples to a bulk effective field theory. In other words, in the presence of the string, the total action can be written as  
\begin{align}\label{AAD}
    S_\text{total}=S_\text{Bulk}+S_\text{String},
\end{align}
where $S_\text{String}$ is a 2d action defined on the locus of the string and which can depend on the values of the spacetime fields in the vicinity of the string. Even the critical string, that is quantized in string theory, fits this description since the critical string has a two-dimensional bosonic action of the form 
\begin{align}\label{critact}
    S_\text{Critical}=-\cT \int \sqrt{-g_\text{ind}}-\frac{1}{4\pi}\int \sqrt{-g_\text{ind}}\mathcal{R}_\text{ind}\phi+i\int B
\end{align}
where $\sqrt{g_\text{ind}}$ and $\mathcal{R}_\text{ind}$ are, respectively, the induced metric on the string worldsheet and its corresponding scalar curvature in the string frame, and $B_{\mu\nu}$ and $\phi$ are spacetime gauge potentials and the dilaton. The difference between the critical string and an effective string is the regime of validity of this action. Suppose the initial state for the string is a closed configuration. Due to its non-zero tension, the string wants to shrink to zero size such that this configuration is unstable. As the string shrinks, the extrinsic curvature of the string, which is inversely related to the size of the string, increases. Since the 2d action of the string depends on the embedding of the string in the spacetime, it can contain a series of higher-order terms in the extrinsic curvature. In fact, as we will discuss below, this is typically required and the series is controlled by the string tension. With that in mind, the higher-order corrections of the string action become important once the size of the string reaches the inverse square root of the tension, i.e. the string length. However, if we were to quantize the string, the average size of the string would be exactly of this order. This is the reason why quantizing a generic string is difficult and why the critical string in string theory is special. In string theory, we assume that we know the full action of the string at weak coupling which allows us to quantize the string in small, closed configurations.

Nonetheless, if one wants to study the string in configurations with small extrinsic curvature, the effective action is sufficient and very useful. Let us mention two examples of such configurations. 
\begin{itemize}
    \item \bf Long string: \normalfont Consider a string in a planar configuration in spacetime. The extrinsic curvature on the worldsheet is zero. Therefore, the use of a finite-term effective action is justified. 
    \item \bf Wrapped string:  \normalfont In the presence of a large extra dimension, one can consider a string that wraps the extra dimension. If the string coupling is small (such that the string does not break up into smaller pieces), the topology of the circle makes this string configuration stable and does not allow it to shrink to zero size. Moreover, if the size of the circle is large, the extrinsic curvature of the string will be very small, allowing the use of a finite-term effective action. 
\end{itemize}
In the case of a fundamental critical string, the worldsheet theory is conformal. Therefore, the worldsheet theory is known at arbitrarily high energies or extrinsic curvatures. This fact is crucial for a successful quantization of the critical string. For an effective string, the theory is not necessarily conformal. However, the worldsheet theory flows to a CFT in the infrared, i.e. for long wavelengths. Therefore, we can assume that in the regime of the validity of the effective string theory (which is long wavelengths) the theory is almost conformal.

Let us focus on the worldsheet action in equation \eqref{AAD}. The full action is model-dependent but there is a universal term imposed by the tension of the string---the Nambu--Goto term,
\begin{equation}
 S_\text{String}=-\cT\int d^2\sigma \sqrt{-g_\text{ind}}+\hdots\,,
\end{equation}
where $g_\text{ind}$ is the 2d metric induced on the string worldsheet 
\begin{align}
    (g_{\rm ind})_{ij}=\frac{\partial X^\mu}{\partial \sigma^i}\frac{\partial X^\nu}{\partial \sigma^j}g_{\mu\nu}\,.
\end{align}This action can be expressed in different coordinate systems on the worldsheet due to the diffeomorphism gauge symmetry. We can go to the light-cone gauge ($\sigma^+,\sigma^-$) to partially fix the gauge  
\begin{align}
    (g_{\rm{ind}})_{++}=(g_{\rm{ind}})_{--}=0\,.
\end{align}
In this gauge, the string action takes the form 
\begin{align}
    S=2\cT\int d\sigma^+d\sigma^-\partial_+X^\mu\partial_-X_\mu+\hdots\,,
\end{align}
and the remnant reparametrization symmetry is given by 
\begin{align}
&\sigma^+\rightarrow f^+(\sigma^+)\,,\nonumber\\
&\sigma^-\rightarrow f^-(\sigma^-)\,.
\end{align}
 Polchinski and Strominger found the appropriate term induced by the Fadeev--Popov determinant which ensures the theory's invariance under the above reparametrization at the quantum level \cite{Polchinski:1991ax}. The action receives a series of corrections that can be organized in terms of the extrinsic curvature of the worldsheet in units of the string tension $\cT$. The leading contribution takes the form
\begin{align}\label{PSA}
    S_{\rm PS}=\int d\sigma^+d\sigma^- 2\cT\partial_+X^\mu\partial_-X_\mu+\frac{d-26}{4\pi}\frac{(\partial^2_+X^\mu\partial_-X_\mu)(\partial_+X^\nu\partial^2_-X_\nu)}{(\partial_+X^\rho\partial_-X_\rho)^2}+\mathcal{O}\left((\cT L^2)^{-1}\right)+\hdots\,,
\end{align}
where $d$ is the dimension of the spacetime and $L\sim\partial X$ is the order of magnitude of the length scale associated with the worldsheet curvature. Note that the above action closely resembles Polyakov's non-critical string action \cite{Polyakov:1981rd}. The action \eqref{PSA} can be transformed into the Polyakov action by using the first-order on-shell equation $\partial_+\partial_-X^\mu=0$ in addition to replacing $\partial_+X^\mu\partial_-X_\mu$ with $\exp(\phi)$.
\begin{align}
        S_\text{Polyakov}=\int d\sigma^+d\sigma^- \cT\partial_+X^\mu\partial_-X_\mu+\frac{d-26}{4\pi}\partial_+\phi\partial_-\phi.
\end{align}
In this theory $\phi$ is an additional scalar. This is due to the fact that for the Polyakov string, the worldsheet metric is auxiliary and $\phi$ is the Goldstone mode of the conformal symmetry on the worldsheet. For the Polchinski--Strominger string, however, everything is formulated in terms of the physical induced 2d metric. 

Note that the second term in \eqref{PSA} is the extrinsic curvature of the string expressed in terms of the spacetime coordinates. The coefficient of the extrinsic curvature can change depending on the rest of the 2d action. For example, in the presence of worldsheet fermions in the neglected ($\hdots$) part of the action, the coefficient changes. However, we can draw the following general conclusion: If the extrinsic curvature is much smaller than the string tension, the leading contribution to the effective action of a weakly coupled string comes from the Nambu--Goto area term.

The weak coupling condition is important to ensure that the self-interaction of the string due to its coupling with bulk modes can be ignored. For example, if the string couples to a 2-form gauge potential as in \eqref{critact}, it will source a $B$-filed background. If the gauge coupling of the $B$-field is significant, the interaction of the string with the $B$-field sourced by the other part of it would not be negligible. Therefore, it is only for weakly coupled strings for which we can ignore their self-interactions through the bulk.

\section{Gross--Mende saddles}\label{GMSA}

In this section we review the Gross--Mende saddles \cite{Gross:1987ar}, of the Polyakov action which play an important role in the contribution of strings to the high-energy scattering amplitudes. Since the Polyakov action and the Nambu--Goto action coincide at the classical level, and the saddle point is a classical configuration, finding a saddle of the Polyakov action also amounts to finding a saddle of the Nambu--Goto action. We will elaborate on this point after reviewing the saddles at the end of the Appendix. 

Let us start by summarizing some special covering spaces of the sphere which are discussed at length in \cite{Hamidi:1986vh} and were used by Gross and Mende \cite{Gross:1987ar} to find worldsheet saddles. Suppose $\{z_1, ..., z_L\}\subset \mathbb{C}$ are $L$ points on this sphere. Then 
 \begin{equation}\label{Ncover}
    y^N=\prod_{i=1}^L (z-z_i)^{L_i}
\end{equation}  
describes a covering of the sphere embedded in $(\mathbb{C}\cup\infty)^2$  provided the integers $L_{1\leq i\leq L}$ satisfy
\begin{equation}\label{IRC}
    \sum_{i=1}^L L_i\equiv 0\mod N\,.
\end{equation}
Since the Polyakov action is invariant under Weyl rescaling, we will work with the flat coordinates $ds^2=dzd\bar z$ that cover all but one point on $\mathbb{CP}^1$. For $z\notin \{z_i|1\leq i\leq N\}$, the above polynomial equation has $N$ complex solutions for $y$. Therefore the space described by \eqref{Ncover} is an $N$-cover of the sphere with $L$ branch points at $z=z_i$ of order $N-1$ (assuming $L_i$'s and $N$ are coprime). The genus of this space is given by 
\begin{align}
    G=\frac{(N-1)(L-2)}{2}\,.
\end{align}
One can verify this using the Gauss-Bonnet theorem. The Euler characteristic of this punctured surface has two contributions, one from the integral of the Ricci scalar over the interior of the surface, and the other from the contribution of the branch points which have angular deficits. Therefore, we find
\begin{align}
    \chi= -\sum_{\text{Punctures}} (\text{Ord}_i+1)+\frac{1}{4\pi}\int_\text{interior} \sqrt{g}\mathcal{R}\,.
\end{align}
Here, $\text{Ord}_i$ represents the order of the branch point at $z=z_i$. The contribution of the first terms is $-LN$ given that there are $L$ branch points each with degree $N-1$. Since the space is an $N$-cover of the sphere, the contribution from integrating the 2d Ricci scalar $N$-times over the smooth component of the surface is $N$ times that of the sphere. The contribution of the second term is $N\cdot \chi_\text{sphere}=2N$. In total this leads to
\begin{equation}
    \chi= -L(N-1)+2N-L.
\end{equation}
To find the genus associated with this surface in the perturbative expansion of string amplitudes, we use 
\begin{equation}
    2-2G-L=\chi =-(N-1)L+2N-L,
\end{equation}
which results in 
\begin{equation}
    G=\frac{(N-1)(L-2)}{2}.
\end{equation}
In the main text we are interested in computing $2\rightarrow 2$ amplitudes describing scattering processes with four external legs. Since external legs correspond to branch points we are mainly interested in covering spaces with $L=4$ branch points. For these covering spaces we have
\begin{equation}\label{GNR}
    G_{2\rightarrow 2}(N)=N-1.
\end{equation}

Having reviewed the covering spaces of the sphere, we are ready to describe the Gross--Mende saddles. Suppose we are interested in a string amplitude involving 4 external legs. Then the amplitude for the $2\rightarrow 2$ process can be determined via the Polyakov path integral in the Euclidean signature, which takes the form 
\begin{equation}
    \mathcal{A}=\sum_\text{topologies}\int \prod_a \left(dt^a f(t^a)\right)\,\mathcal{D}X^\mu\, e^{-S_\text{Polyakov}}\prod_{i=1}^4 \left(d z_i \right) :e^{ip_i\cdot X(z_i)}:\mathcal{V}_i(z_i)\,.
\end{equation}
Let us explain the ingredients in the above path integral:
\begin{itemize}
    \item The operators $\cV_i$ are the vertex operators that identify the species of the external legs. Note that for every pair of conformal Killing vectors, there is a vertex operator $\mathcal{V}_i$ that contains a $c$-field ghost insertions multiplied by $\delta(z-z')$ that fixes the location of $z_i$. 
    \item The vectors $p_i$ are the momenta of the external legs and $::$ represents worldsheet normal ordering to regulate the worldsheet OPEs. 
    \item The integration variables $t^a$ are the moduli of the Riemann surface.
    \item The integration variables $z_i$ are complex worldsheet coordinates corresponding to the location of the insertions.
    \item  The factor $f(t^a)$ is the integration measure over the moduli space which is determined by the Fadeev--Popov determinant and $b$-field ghost insertions. 
\end{itemize}
    Note that here we neglected the fermionic worldsheet fields for simplicity as their inclusion does not affect our final conclusions. At high energies ($p_i\gg M_s$), the behavior of this path integral is not sensitive to the choice of the species of the in/out-going particles. Therefore, the amplitude can be approximated by the 4-point function of the $:e^{ip_i\cdot X(z_i)}:$ operators which is given by
\begin{equation}
     \mathcal{A}\sim \delta^d\left(\sum_i p_i\right) \int \prod_{j} d z_j \exp\left(\frac{1}{2}\sum_{1\leq i<j\leq 4}^4 p_i\cdot p_j G(z_i,z_j)\right)\,,
\end{equation}
where $G(z_i,z_j)$ is the propagator on the surface. Here we use $\cT=1/(2\pi \alpha')$ and our convention for the signature of the spacetime metric is $(-,+,\hdots,+)$. Gross and Mende approximated this integral using the saddle point approximation. For that, we need to find the saddle point of $\mathcal{E}=\frac{1}{2}\sum_{i,j}(-p_i\cdot p_j) G(z_i,z_j)$. Fortunately, this is a familiar problem since $G(z_i,z_j)$ is the solution to $$\partial\bar\partial G(z,z_0)=-\left(\frac{\alpha'\pi}{2}\right)\delta^2(z-z_0)\,,$$ which also describes the electrostatic potential of a charged particle with unit charge at $z=z_0$ in two Euclidean dimensions. Therefore, $\mathcal{E}$ is the electrostatic energy of a configuration of particles with charges $p_i$ at points $z_i$.

The mimimum energy configuration depends on the topology of the saddle. Let us start with the simplest case of the sphere. On the sphere, the propagator is $G(z_i,z_j)=-(\alpha'/4)\log|z_i-z_j|^2$ and the energy is given by 
\begin{align}
    \mathcal{E}=\frac{\alpha'}{4}\sum_{1\leq i<j\leq4}p_i\cdot p_j \log|z_i-z_j|\,,
\end{align} 
which can be expressed in terms of the cross ratio $$\lambda=\frac{(z_1-z_3)(z_2-z_4)}{(z_1-z_2)(z_3-z_4)}\,,$$ as
\begin{equation}
   \mathcal{E}=-\frac{\alpha'}{8}(t\log|\lambda|+u\log|1-\lambda|)\,,
\end{equation}
where $s$, $t$, and $u$ are the usual Mandelstam variables (see~\eqref{mandelstam}). In writing the above equation, we have used $s\simeq -t-u$ which is a good approximation at high energies $E\gg M_s$ where the external legs can be treated as massless relativistic particles. The extremum of the above action is reached for 
\begin{align}\label{saddle}
    \lambda_{\rm saddle}=-t/s\simeq 1+u/s\,.
\end{align}
For this critical $\lambda$, the amplitude is given by 
\begin{align}\label{GMLS}
    \mathcal{A}_\text{tree}\sim \exp\left(-\frac{\alpha'}{8}(s\log s+u\log u+t\log t)\right)\,,
\end{align}
which agrees with the high-energy behavior of the Virasoro--Shapiro amplitude. Note that there are also polynomial factors multiplying the exponential term that result from integrating around the saddle \eqref{saddle}.

To study surfaces with higher genus, we can repeat the same analysis for the covering spaces of the sphere which we discussed earlier. Consider an $N$-cover of the sphere. The energy function $\mathcal{E}$ is suppressed by a factor $1/N$. This is because on every sheet the \textit{electric charges}, and the corresponding \textit{electric field} $E$, are divided by $N$. On the other hand, the energy $\mathcal{E}$, is proportional to the integral of $E^2$ over $N$ sheets which leads to a $N\times (1/N^2)=1/N$ suppression compared to the tree-level diagram. The rest of the analysis gives the same cross-ratio $\lambda$ for the positions $z_i$. Therefore, Gross and Mende found
\begin{align}\label{OMEGM}
    \mathcal{A}_{N}\sim g_s^{2G-2}\exp\left(-\frac{\alpha'}{8N}(s\log s+u\log u+t\log t)\right)\,,
\end{align}
where $g_s$ is the string coupling. Using \eqref{GNR}, we can express $G$ in terms of $N$. 
\begin{align}
    \mathcal{A}_{N}\sim g_s^{2N-4}\exp\left(-\frac{\alpha'}{8N}(s\log s+u\log u+t\log t)\right)\,,
\end{align}
Every $N$ corresponds to a different saddle. However, one can see that the amplitude is maximized for 
\begin{equation}\label{dominantsaddle}
    N_{\rm max}\simeq \sqrt{\frac{\alpha'\left(s\log s+u\log u+t\log t\right)}{16\left(-\log g_s\right)}}\,,
\end{equation}
for which we have 
\begin{equation}
    \mathcal{A}\sim \exp(-\sqrt{\left(-\log(g_s)\right)\alpha'\left(s\log s+u\log u+t\log t\right)})\,,
\end{equation}
up to multiplication by factors polynomial in energy. Therefore, we conclude that the dominant contribution from the Gross-Mende saddles goes like 
\begin{equation}
        \log(\mathcal{A})\sim -\sqrt{s}\cdot f(s,\theta)\,,
\end{equation}
where $\theta$ is the scattering angle and $f(s,\theta)$ is a function with subpolynomial (logarithmic) growth/decay in $\sqrt s$. Note that the polynomial prefactors, which we neglected in \eqref{OMEGM}, contribute to $f(s,\theta)$

It is also instructive to review the embedding of these saddles in spacetime. Around each external leg, the position is given by $X^\mu\simeq \frac{ip_j^\mu}{4\pi \cT}\log|z-z_j|$, where $\cT=1/(2\pi\alpha')$ is the string tension. Therefore, we have
\begin{equation}\label{GME}
    X^\mu=\sum_{j=1}^4\frac{ip_j^\mu}{4\pi N \cT}\log|z-z_j|+\mathcal{O}\left(\frac{\sqrt{\cT}}{s}\right)\,.
\end{equation}
From the above expression, one can see that the length scale associate with the extrinsic curvature ($L\sim \partial X^\mu$) behaves like $L\sim E/(N\cT)$ where $E$ is the center of mass energy. For the dominant saddle, using \eqref{dominantsaddle} this length scale evaluates to $L\sim l_s$ up to multiplicative logarithmic corrections in energy. Note that for $1\ll N\sim N_{\rm max}(E/M_s)^{-\epsilon}\ll N_{\rm max}$, where $\epsilon>0$ and $E\gg M_s$, we have
\begin{equation}
    L\gg l_s~~\text{and}~~\log(\mathcal{A})\sim -\frac{E}{\sqrt{\cT}}\widetilde{\mathcal{O}}\left(L\sqrt{\cT}\right)\,.
\end{equation}
In the above discussion, we used the Polyakov action to find the saddle. Saddles satisfy the classical equations of motion, and both Polyakov and Nambu--Goto action lead to the following equation of motion for the embedding of the worldsheet 
\begin{align}
    \Box X^\mu=0\,,
\end{align}
where $\Box$ is the d'Alembertian with respect to the induced metric $h_{ij}=\partial_i X^\mu\partial_j X_\mu$. However, one might be concerned whether the boundary conditions imposed at the external legs are the same in the two pictures. For each external leg, we have chosen a spacetime momentum which is the conserved charge under translations in $X^\mu$. In the Polyakov action this conserved quantity is given by 
\begin{align}
    p^\mu=\frac{\cT}{2}\int d\sigma \sqrt{h}\,h^{0i}{\partial_i}X^\mu\,,
\end{align}
while in the Nambu--Goto action, this conserved charge is 
\begin{align}
    p^\mu=\frac{\cT}{2}\int d\sigma \sqrt{g_\text{ind}}\,g_\text{ind}^{0i}{\partial_i}X^\mu\,,
\end{align}
where $g_\text{ind}$ is the induced metric on the worldsheet. Therefore, after plugging in the equations of motion $h\propto g_\text{ind}$, we see that the two definitions of momenta match. In other words, the embedding \eqref{GME} which is a Polyakov saddle, is also a solution to the Nambu-Goto equations of motion with external legs that carry spacetime momenta $p_i^\mu$. Therefore, the same embeddings are also saddles of the Nambu--Goto action.

\section{Amplitude toolbox}\label{app:amplitude}

In this Appendix we review some results about scattering amplitudes and how they apply to high-energy amplitudes in quantum gravity. We focus on $2\rightarrow 2$ scattering. 

Let us start by defining our conventions and notations. The in-going spacetime momenta are $k_1$ and $k_1$ and the out-going spacetime momenta are $-k_3$ and $-k_4$. Our convention for the signature of the spacetime metric is $(-,+,\hdots,+)$. The conservation of energy-momentum implies $\sum_ik_i^\mu=0$. The Mandelstam variables are defined as 
\begin{equation}\label{mandelstam}
\begin{aligned}
    s&=-(k_1+k_2)^\mu(k_1+k_2)_\mu\,,\\
    t&=-(k_1+k_3)^\mu(k_1+k_3)_\mu\,,\\
    u&=-(k_1+k_4)^\mu(k_1+k_4)_\mu\,.
\end{aligned}
\end{equation}
The scattering angle $\theta$ can be expressed in terms of $k_1$ and $k_3$ as 
\begin{equation}
    \cos(\theta)=\frac{\sum_{\alpha \neq0} k_1^\alpha{k_3}_\alpha}{\sqrt{(\sum_{\beta\neq 0}k_1^\beta{k_1}_\beta)(\sum_{\gamma\neq 0}k_3^\gamma{k_3}_\gamma)}}\,.
\end{equation}
Suppose the in-going and out-going particles are on-shell and have physical mass $m$. We can express the scattering angle in terms of the Mandelstam variables as
\begin{align}
    \cos(\theta)=1+\frac{2t}{s-4m^2}\,.
\end{align}
Since $|\cos(\theta)|\leq1$, the variable $t$ is bounded to lie in the interval $[-s+4m^2,0]$. This regime is referred to as the physical regime for $t$. 

\subsection{Regge behavior}\label{RBA}

An interesting limit in the scattering processes is obtained by keeping $t$ fixed while $s$ is taken to infinity. In this limit, the scattering angle $\theta$ goes to 0. Therefore, this limit corresponds to two very high energy particles that barely change each other's trajectory while exchanging a fixed amount of spacetime momentum $t$. The dependence of the t-channel amplitude on $s$ for the exchange of a spin-$L$ particle can be determined in the following way. A spin-$L$ particle is represented by a traceless symmetric field with $L$ indices. Suppose a vertex has two scalar legs and one spin-$L$ leg. Due to Lorentz invariance, the vertex must contain $L$ derivatives. Therefore, each one of the two vertices in the t-channel yields a contribution proportional to $(i\sqrt{s})^L$ in the high energy limit. Therefore, the overall amplitude is proportional to $(-s)^L$. 

The contribution of the exchange of such a particle to the t-channel amplitude depends on $t$ as well. As $\sqrt{{-t}}$ approaches the mass of a particle, the amplitude diverges. However, for a fixed value of $t$, at high center of mass energies, the contribution of this particle exchange grows polynomially as $(-s)^L$. Note that this is not the full t-channel amplitude since there could be more contributions to the t-channel. However, for fixed $t$, the number of weakly coupled particles that can participate in the t-channel exchange will be finite. Therefore, the polynomial growth of the amplitude is expected to continue as long as we are at a generic value of $t$ which does not coincide with the mass of the exchanged particles. This behavior is called the Regge behavior. To be more precise, for every fixed $t$ and $s\gg |t|$  we can parameterize the amplitude as in \cite{Caron-Huot:2016icg} in terms of a continuous function $j(t)$ and a meromorphic function $F(t)$ as
\begin{align}\label{reggebehavior}
    \mathcal{A}\simeq F(t)(-s)^{j(t)}\,.
\end{align}
In tree-level string theory $j(t)$ is a linear function in $t$. In \cite{Caron-Huot:2016icg}, the authors showed that in the presence of higher-spin weakly coupled (tree-level) particles, linearity of $j(t)$ at high energies is universal. As we discussed above, a single spin-$L$ particle with mass $m$ will result in a non-negative integer value $j(t)=L$ and a pole for $F(t)$ at $-m^2$. Therefore, we expect $j(t)$ to be integer at poles of $F(t)$ and vice versa (see A.2 in \cite{Caron-Huot:2016icg} for an argument).

\subsection{Locality bound}\label{app:locality}

In this subsection we briefly review the locality bounds on amplitudes and discuss how (and if) they apply to quantum gravity.  

The most well-known bound on high-energy scattering amplitudes based on locality is the Cerulus--Martin bound \cite{Cerulus:1964cjb} which is a lower bound for the amplitude of a hard scattering process at high energies. The Cerulus--Martin bound is obtained under certain assumptions on the amplitude. In particular, the amplitude is assumed to have the following three properties: 
\begin{enumerate}
    \item The amplitude is analytic as a function of $s$ and $z=\cos(\theta)$ in a complicated sub-region of the cut $z$-plane which is described in \cite{Cerulus:1964cjb} (see section 3.2 in \cite{Buoninfante:2023dyd} for a more detailed explanation).
    \item The amplitude is polynomially bounded at high energies for a fixed range of scattering angles, $z\in [-a,a]$ dor some fixed $a$, 
    \begin{align}
        |\mathcal{A}(s,z)|<A\left(\frac{s}{s_0}\right)^N,
    \end{align}
    where $\sqrt{s_0}$ is some reference energy and $A$ and $N$ are constants. This assumption is often understood to follow from locality since amplitudes that are reproduced by a local effective field theory with finitely many terms are polynomially bounded. 
    \item The forward scattering amplitude is polynomially bounded from below
    \begin{align}
        |\mathcal{A}(s,1)|>\frac{B}{(\frac{s}{s_0})^\beta}\,,
    \end{align}
    for some positive constants $B$ and $\beta$ at high energies (i.e. $s\gg s_0$). 
\end{enumerate}
Given an amplitude that satisfied these three assumption for a fixed range of scattering angles $z\in [-a,a]$, the Cerulus--Martin bound states that the amplitude is bounded as
\begin{align}
    |\mathcal{A}(s,z)|>\mathcal{N}(s)\,e^{-f(a)\sqrt{s}\log(s/s_0)}\,,
\end{align}
where $\mathcal{N}(s)$ is a subdominant function of $s$ and $f(a)$ is independent from $s$.\\

Given this bound, we can now ask whether and under what conditions we expect it to be satisfied in theories of quantum gravity. We first notice that the Cerulus--Martin bound is violated in tree-level string theory since tree-level amplitudes violate polynomial boundedness \cite{Gross:1987kza}. Therefore, the tree-level string theory does not exhibit the local properties of an effective field theory. As a consequence high-energy tree-level amplitudes in string theory only satisfy a bound that is weaker than the Cerulus--Martin bound (cf. \eqref{GMLS})
\begin{align}
    |\mathcal{A}_{\rm tree}(s,z)|>\mathcal{N}(s)e^{-f(a)\,s\log(s/s_0)}\,.
\end{align}
In \cite{Buoninfante:2023dyd} the Cerulus-Martin bound was generalized to exponentially bounded amplitudes and the application of the generalized lower bound to tree-level string theory yields the lower bound
\begin{align}
        |\mathcal{A}_{\rm tree}(s,z)|>\mathcal{N}(s)e^{-f(a)s^{3/2}}\,,
\end{align}
which is satisfied. Interestingly, when the higher-genus contributions from the Gross--Mende saddles, discussed in appendix~\ref{GMSA}, are included, the overall amplitude can be resummed to~\cite{Mende:1989wt}
\begin{align}
    |\mathcal{A}(s,z)|>\mathcal{N}(s)e^{-f(a)\sqrt{s}\log(s/s_0)^{q}},
\end{align}
where $q$ is a negative number. The above amplitude satisfies the Cerulus--Martin bound. It thus seems that including higher-genus corrections restores some notion of locality. Nonetheless, the exact connection to locality is not well-understood (see section 6.4 in \cite{Buoninfante:2023dyd}). 

Still, under certain assumptions, we can motivate the Cerulus--Martin bound for high-energy amplitudes in theories of gravity. However, instead of showing the bound in the fixed angle scattering, we first argue that it holds for any fixed impact parameter. The inequality then follows for the fixed angle scattering amplitude since it can be estimated using a saddle point approximation to be of the order of the fixed impact parameter that forms a saddle~\cite{Amati:1988tn}. 

In section \ref{sec:Hagedorn}, we argued that the density of one-particle states at energies above the quantum gravity cutoff $\Lambda$ must grow exponentially in energy. As reviewed in section~\ref{OPS}, we assume that the density of states is related to the high-energy/low-impact parameter scattering amplitude as $\log \rho(\sqrt{s})\sim-\log(|\mathcal{A}(s,b)|^2)$. The intuition behind this assumption is that at high-energies the in-going particles form bound states/saddles that exhibit thermal properties with $e^S$ microstates such that $S\sim \log(\rho)$. Therefore, the probability of having that thermal state decay into a 2-particle  state is $e^{S}$. This assumption is correct in string theory \cite{Bedroya:2022twb} even though a bottom-up argument justifying this assumption is missing. Based on this assumption, we find that the high-energy amplitude for energies $\Lambda\ll \sqrt{s}<M_\text{min}$ satisfies the Cerulus--Martin bound as
\begin{align}
    \log(|\mathcal{A}(s,b)|^2)\gtrsim - \log \rho(\sqrt{s})=-f(s)\sqrt{s},
\end{align}
where $f(s)$ has logarithmic growth. The above inequality is saturated at low impact parameters, but not at large impact parameters where less intermediate channels are available.

\subsection{Unphysical amplitudes (\texorpdfstring{$s,t\gg \Lambda$}{s,t>>Lambda})}\label{app:unphysical}

As we explained in the beginning of this Appendix, the Mandelstam variable $t$ can only take negative values for on-shell particles. However, in this section we are interested in the analytic continuation of the amplitude to the unphysical regime in which $t$ takes large, positive values. As we will explain, unlike the physical fixed angle amplitudes, the amplitude in this regime is very sensitive to the aforementioned suppression at small impact parameters. 

Let us start by explaining the relationship between scattering amplitudes in terms of the impact parameter and Mandelstam variables. The impact parameter $b$ is the relative position between the two wave-packets while $q=p_1+p_3$ is the momentum exchange between the two particles. The component, $K_{\perp}$, of the spatial part of $q$ perpendicular to $p_1$ is conjugate to $b$ such that the amplitude $\mathcal{A}(s,b)$ can be obtained from $\mathcal{A}(s,t)$ via
\begin{align}
    \mathcal{A}(s,b)=\int d^{d-2}K_\perp e^{i\vec K_\perp\cdot \vec b}\mathcal{A}(s,t)\,.
\end{align}
Since $t$ and $b$ are conjugate to each other they cannot both be known. As explained in the previous subsection, at energies $\sqrt{s}\gg \Lambda$, we expect the amplitude $\mathcal{A}(s,b)$ to be exponentially suppressed at sufficiently small impact parameters. For example, if the center of mass energy $\sqrt{s}$ is higher than the black hole formation threshold $m_\text{min}$, the amplitude satisfies
\begin{align}
    \log(\mathcal{A}(s,b\ll r_s(M)))\sim -S(M)\,,
\end{align} 
where $r_s(M)$ and $S(M)$ are respectively the radius and entropy of a black hole with mass $M$. This exponential suppression cannot be easily seen in $\mathcal{A}(s,t)$ because $\mathcal{A}(s,t)$ is the Fourier transform of $\mathcal{A}(s,b)$ and receives contributions from all impact parameter. Therefore, the exponentially small contribution of low-impact parameters is overshadowed by the polynomially large contributions of the amplitudes at high impact parameters. On the other hand, the exponential suppression at small $b$ leaves an exponential phase imprint in the amplitude $\sim \exp(2\sqrt{t}\,r_s(\sqrt{s}))$. This term becomes dominant in the unphysical regime $t>0$ (see \cite{Giddings:2007qq} for details) such that this regime is a good probe for the exponential suppression of the amplitudes at small impact parameters. 

More generally, the same argument also applies to non-black hole bound states as we show in the following. Suppose the amplitude $\mathcal{A}(s,b)$ is exponentially suppressed for impact parameters much smaller than some critical value $b_c(\sqrt{s})$. Further assume that it is polynomial for impact parameters $b\gg b_c$. By modeling the scattering process with a black disk that blocks low-impact parameter amplitudes one can find that the amplitude in the unphysical regime behaves as 
\begin{align}\label{UPA}
    \mathcal{A}\left(s,t\gg b_c(\sqrt{s})^{-2}\right)=\mathcal{N}(s,t)\exp(\sqrt{t}\,\mathcal{O}(b_c(\sqrt{s}))\,, 
\end{align}
where $\cN(s,t)$ is a subdominant function. This statement was proven in \cite{Giddings:2007qq} for black holes but the same argument applies to any amplitude which is exponentially suppressed at low impact parameters. The reason is that the diffraction off of the edge of the black disc will lead to contributions proportional to $\sim \sin(\mathcal{O}(E\theta\,b_c(\sqrt{s})))$ to the amplitude. If we express the scattering angle in terms of Mandelstam variables $-t/s\simeq \theta^2/4$ and analytically continue to positive $t$ we find \eqref{UPA}. 

In section~\ref{app:locality} we argued that the low-impact parameter amplitude in quantum gravity should be exponentially suppressed based on the identification between the density of one-particle states and the scattering amplitude. Assuming that this exponential suppression is realized for $b\ll b_c(\sqrt{s})$ we find that the unphysical high-energy amplitude below the black hole threshold is given by \eqref{UPA}. Moreover, we know that this amplitude must match with that of the black hole at the black hole threshold $\sqrt{s}\sim M_\text{min}$. Above the black hole threshold, $b_c$ is given by the radius of the black hole. Therefore, we find 
\begin{align}
    b_c(M_\text{min})=r_{\min} = \Lambda_{\rm min}^{-1}\,.
\end{align}
While this relation determines $b_c(\sqrt{s})$ at energies corresponding to the black hole threshold, it does not yet fix the energy dependence of $b_c$ at energies below $M_{\rm min}$. We can provide a bound on the energy dependence of $b_c$ below the black hole threshold by using the Regge behavior reviewed in section~\ref{RBA} which is expected to hold below the black hole threshold. From \eqref{UPA}, we conclude that if $b_c(\sqrt{s})$ depended polynomially on $\sqrt{s}$ for $\sqrt{s}<M_{\rm min}$, the off-shell amplitude $\mathcal{A}(s, t>0)$ would have an exponential dependence on $s$. By \eqref{reggebehavior} such an exponential dependence is, however, inconsistent with the Regge behavior which forces $b_c(\sqrt{s})$ to have at best a logarithmic dependence on $\sqrt{s}$ at high energies above the cutoff $\Lambda_{\rm min}$. Still, the Regge behavior is motivated for a perturbative description of the amplitudes which we expect to be violated at the black hole threshold $\sqrt{s}=M_{\rm min}$. Therefore, we find that for energies $\Lambda_{\rm min} \ll \sqrt{s}\ll M_{\rm min}$,
\begin{align}\label{ipb}
b_c(\sqrt{s})^{-1}=\widetilde{\mathcal{O}}\left(\Lambda_{\rm min}\right)\,.
\end{align}
Plugging equation \eqref{ipb} back into \eqref{UPA} leads to 
\begin{align}\label{EAEFT}
    \mathcal{A}(s,t\gg \Lambda_{\rm min}^2)\sim \mathcal{N}(s,t)\exp\left[\widetilde{\mathcal{O}}(\sqrt{t}/\Lambda_{\rm min})\right]\,.
\end{align}
This result for the amplitude is the main ingredient for our argument in favor of Proposition~\ref{prop:existenceofEFT} in section~\ref{sec:ExistenceEFT}.

\bibliographystyle{utphys}
\bibliography{papers}

\end{document}